\documentclass[12pt]{article}
\usepackage{amssymb,amsmath}

\newcommand{\<}{\langle}
\renewcommand{\>}{\rangle}

\newcommand{\publititle}[8]
{ \vspace*{-3cm}
  \begin{flushright} #1 \\ {\tt #2} \end{flushright}
  \vfill
  \begin{center}{\LARGE
    \bfseries #3}\end{center}
  \vskip 8mm
  \begin{center}{\large #4}\end{center}
  \begin{center}{\normalsize #5}\end{center}
  \vskip 8mm
  \nopagebreak
  \noindent #6
  \vfill
  \hrule width 6.7cm \vskip.1mm
  \begin{flushleft} #7
  \end{flushleft}
  {\small #8}
  \thispagestyle{empty}
  \clearpage
}

\begin{document}

\publititle{ ${}$ \\ UCB-PTH-02/35 \\ LBNL-51253 \\ hep-th/0208150}{}
{On the Large N Limit \\ of Conformal Field Theory}
{M.~B. Halpern$^a$}
{Department of Physics, \\
University of California, 
Berkeley, California 94720, USA\\
{\it and}
Theoretical Physics Group,
Lawrence Berkeley National Laboratory\\
University of California,
Berkeley, California 94720, USA \\ \hskip 1cm \\
\today
\\[2mm]}{Following recent advances in large $N$ matrix mechanics, I discuss here the
free (Cuntz) algebraic formulation of the large $N$ limit of two-dimensional conformal field 
theories of chiral
adjoint fermions and bosons.  One of the central results is a new 
{\it affine free algebra}
which describes a large $N$ limit of $\mathfrak{su}(N)$ affine Lie algebra. 
Other results include
the associated {\it free-algebraic partition functions and characters}, a 
free-algebraic coset
construction, free-algebraic construction of $\mathfrak{osp}(1|2)$, 
{\it free-algebraic vertex operator
constructions} in the large $N$ Bose systems and a provocative new 
free-algebraic
factorization of the ordinary Koba--Nielsen factor.}{$^a${\tt 
halpern@physics.berkeley.edu}
}

%%\title{{\bf On the Large ${\mathbf N}$ Limit \\ of Conformal Field Theory}}%
%%\author{M.~B. Halpern\footnote{halpern@physics.berkeley.edu}\\
%%Department of Physics\\
%%University of California\\
%%Berkeley, California 94720, USA\\
%%and\\
%%Theoretical Physics Group\\
%%Lawrence Berkeley National Laboratory\\
%%University of California\\
%%Berkeley, California 94720, USA}

%%\maketitle

%%\noindent  Following recent advances in large $N$ matrix mechanics, I discuss here the
%%free (Cuntz) algebraic formulation of the large $N$ limit of two-dimensional conformal field 
%%theories of chiral
%%adjoint fermions and bosons.  One of the central results is a new 
%%{\it affine free algebra}
%%which describes a large $N$ limit of $\mathfrak{su}(N)$ affine Lie algebra. 
%%Other results include
%%the associated {\it free-algebraic partition functions and characters}, a 
%%free-algebraic coset
%%construction, free-algebraic construction of $\mathfrak{osp}(1|2)$, 
%%{\it free-algebraic vertex operator
%%constructions} in the large $N$ Bose systems and a provocative new 
%%free-algebraic
%%factorization of the ordinary Koba--Nielsen factor.

%%\clearpage
\tableofcontents

\clearpage
\section{Introduction}
\label{sec1}

{\it Large $N$ matrix mechanics} [1-6] is the adaptation of Heisenberg's matrix mechanics
[7] to study systems of $\mathfrak{su}(N)$ adjoint matter at large $N$.  For both
Hamiltonian [1-4,6] and action [5,6] theories, this approach gives a  
simpler ``reduced"
formulation 
of the {\it adjoint sector} of the large $N$ theory in terms of so-called 
{\it free algebras}. 
The simplest free algebra is the {\it Cuntz algebra}, but many generalizations [4,5] 
have also been found in large $N$ matrix mechanics, including the interacting Cuntz 
algebras and various fermionic and supersymmetric versions of the Cuntz algebra. 
In all cases, the free algebras are associated to  new, hidden conserved 
quantities at large $N$ -- and corresponding simplifications of the large N theory.

Recently, Halpern and Thorn [8]  showed for any matrix light-cone Hamiltonian 
field theory that the large $N$ reduced Hamiltonian of the theory can be expressed 
in terms of the generators of the Cuntz algebra itself, with correspondingly 
enhanced simplifications.  As simple examples of this result, I consider here the large 
$N$ limit of {\it $2$-d} matrix conformal field theory -- which always has such a light-cone 
formulation.

More precisely I will follow an historical development, limiting the discussion 
here to the matrix analogues of the earliest and simplest {\it $2$-d} conformal and 
superconformal field theories -- namely those composed of free chiral adjoint 
world-sheet fermions and bosons. In spite of the apparent simplicity of these 
theories at finite values of $N$, we shall see that their large $N$ limit shows a 
wealth of new phenomena, all succinctly described in terms of free algebras.

For example, the reduced, Cuntz-algebraic description of  large $N$ chiral adjoint 
fermions is studied in Subsec.2.1 and, following the independent discovery of 
affine Lie algebra in physics [9], {\it reduced currents} are constructed
as reduced 
fermion bilinears in Subsec.2.2. The reduced current modes satisfy a new 
{\it affine free algebra} at reduced level $\hat{x}$
\renewcommand{\theequation}{\thesection.\arabic{equation}}
\setcounter{equation}{0}
\begin{equation}
\label{eq1.1}
J(m \ge 0)J(n \le 0) = -J(m+n) + {\hat x}m\delta_{m+n,0}
,\quad m,n \in {\mathbb Z}
\end{equation}
which describes the large $N$ limit of $\mathfrak{su}(N)$ affine Lie algebra in the adjoint
sector. The representation theory of this new algebra is surprisingly rich
because, as we will see, many new affine primary states occur in the large $N$
limit which are not present at any finite value of $N$. 

Other large $N$ results include the corresponding {\it free-algebraic partition
functions and characters} (which exhibit a surprising {\it $2$-d limiting temperature}),
an extended affine free algebra and  a free algebraic coset construction,
as well as a free-algebraic construction of $\mathfrak{osp}(1|2)$. In the last section,
I consider a set of more exotic free-algebraic constructions in the large $N$
Bose systems -- including {\it free-algebraic vertex operators} and free-algebraic
vertex operator constructions of {\it new excitations} at large $N$.  This leads us
finally to a provocative new free-algebraic factorization of the familiar
Koba-Nielsen factor [10] and hence to the question of  new {\it free-algebraic
strings}.

\section{Chiral Adjoint Fermions}
\label{sec2}
It is an historical curiosity that the discovery of  half-integral moded
chiral world-sheet fermions [9] and  integral-moded world sheet fermions [11]
appeared as consecutive articles in the same issue of Physical Review. In the
large $N$ formulations of this section , I will follow the development of Bardakci
and Halpern [9,12], who  used the world-sheet fermions to construct the
so-called {\it dual quark  model}, including the independent discovery of affine Lie algebra
in physics and the first examples of the affine-Sugawara and coset  constructions.
See the Appendix of Ref. [13] for a short history of affine Lie algebra and
associated early developments in conformal field theory.

\subsection{Cuntz-algebraic description of the large $N$ limit}
\label{sec2.1}

To avoid the complexities of fermionic zero modes, I will begin with a
single half-integral moded chiral world-sheet fermion in the adjoint of 
$\mathfrak{u}(N)$
\renewcommand{\theequation}{\thesection.\arabic{subsection}\alph{equation}}
\setcounter{subsection}{1}
\setcounter{equation}{0}
\begin{equation}
\label{eq2.1a}
[H_a(p),H_b(q)]_+ = \delta_{ab}\delta_{p+q,0},\quad H_a(p)^{\dag} = H_a(-p)
\end{equation}
\begin{equation}
\label{eq2.1b}
H_a(p>0) |0.\> = \<.0| H_a(p<0) = 0
\end{equation}
\vspace{-0.2in}
\begin{equation}
\label{eq2.1c}
\<.0|0.\> = 1,\quad a,b = 1\dots N^2,\quad p,q \in {\mathbb Z} + \frac {1}{2}
\end{equation}
which approximates the adjoint of $\mathfrak{su}(N)$ at large $N$.  Systems of many chiral
adjoint fermions will be considered later. The development of this subsection
is an infinite-dimensional generalization of the discussion given for
quantum-mechanical adjoint fermions in Ref.[4].

For fields in the adjoint, the first step is always to introduce matrix fields
\setcounter{subsection}{2}
\setcounter{equation}{0}
\begin{equation}
\label{eq2.2a}
H(p)_{rs} = H_a(p)(T_a)_{rs},\quad H(p)_{rs}{}^{\dag} = H(-p)_{sr}
\end{equation}
\begin{equation}
\label{eq2.2b}
[T_a,T_b] = if_{abc}T_c,\quad T_a{}^{\dag} = T_a,\quad \mbox{Tr}(T_aT_b) = \delta_{ab}
\end{equation}
\begin{equation}
\label{eq2.2c}
(T_a)_{rs}(T_a)_{uv} = \delta_{rv}\delta_{su},\quad r,s,u,v = 1\dots N
\end{equation}
\begin{equation}
\label{eq2.2d}
[H_{rs}(p),H_{uv}(q)]_+ = \delta_{rv}\delta_{su}\delta_{p+q,0}
\end{equation}
\begin{equation}
\label{eq2.2e}
H_{rs}(p>0) |0.\> = \<.0| H_{rs}(p<0) = 0
\end{equation}
where $\{T_a\}$ is the fundamental representation of $\mathfrak{u}(N)$ and I have chosen root
length squared $\alpha^2 =2$ for $\mathfrak{su}(N)$.  In large $N$ matrix mechanics [1-6]
we restrict the Hilbert space to the {\it adjoint sector} at large $N$, which includes
only the singlet ground state and the {\it dominant adjoint eigenstates} $|...\>_{rs}$
which saturate
the traced Wightman functions  $\<.0|\mbox{Tr}(H(p_1)...H(p_n))| 0.\>$ of the large $N$
theory. The large $N$ dynamics is then mapped by a generalization of
Bardakci's [14] reduced matrix elements (Wigner-Eckart for $\mathfrak{su}(N)$) onto
{\it reduced states} and {\it reduced operators} [4]
 \setcounter{subsection}{3}
 \setcounter{equation}{0}
 \begin{equation}
 \label{eq2.3a}
 \left( \frac {H(p)}{\sqrt{N}} \right)_{rs} \rightarrow H(p),{\tilde H}(p)
\end{equation}
\vspace{-0.2in}
\begin{equation}
 \label{eq2.3b}
 \<.0| \mbox{Tr}\left( \frac {H(p_1)}{\sqrt{N}} \dots \frac 
 {H(p_n)}{\sqrt{N}} \right)
 |0.\> = N \<0|H(p_1)\dots H(p_n)|0\>
\end{equation}
where $|0\>$ is the reduced ground state. Note that each reduced operator 
$H$ is accompanied by a reduced tilde partner ${\tilde H}$, and that all $r,s$ indices
are modded out in the reduction procedure.  The relation in (2.3b) is only a sample, 
the full list of reduction formulae being given in Ref. [4]. Next, one finds
from Eq. (2.29) of Ref. [4] that the reduced operators satisfy the
quasi-canonical algebra
\setcounter{subsection}{4}
\setcounter{equation}{0}
\begin{equation}
\label{eq2.4a}
[H(p),{\tilde H}(q)]_+ = |0\>\<0| \delta_{p+q,0},\quad H^{\dag}(p) = 
H(-p),\ {\tilde
H}^{\dag}(p) = {\tilde H}(-p)
\end{equation}
\begin{equation}
\label{eq2.4b}
H(p>0)|0\> = {\tilde H}(p>0)|0\> = \<0| H(p<0) = \<0| {\tilde H}(p<0) = 0
\end{equation}
\begin{equation}
\label{eq2.4c}
({\tilde H}(p<0) - H(p<0))|0\> = \<0| ({\tilde H}(p>0) - H(p>0)) = 0
\end{equation}
\begin{equation}
\label{eq2.4d}
\<0|0\> = 1.
\end{equation}
As discussed in Ref.[4], a basis for a complete set of reduced states
can be constructed by acting on the reduced vacuum with the creation
operators $\{H(p<0)\}$ or equivalently  by acting with the tilde
creation operators $\{{\tilde H}(p<0)\}$.  Finally, the action of
the algebra (2.4) on either basis gives the further free-algebraic relations
\setcounter{subsection}{5}
\setcounter{equation}{0}
\begin{equation}
\label{eq2.5a}
H(p>0)H(q<0) = \delta_{p+q,0},\quad \sum_{p > 0} H(-p)H(p) = 1-|0\>\<0|
\end{equation}
\vspace{-0.2in}
\begin{equation}
\label{eq2.5b}
{\tilde H}(p>0){\tilde H}(q<0) = \delta_{p+q,0},\quad \sum_{p>0} {\tilde 
H}(-p){\tilde H}(p)
= 1 - |0\>\<0|
\end{equation}
which,  together with the relations  in (2.4),  form an infinite-dimensional
version of a so-called {\it symmetric Fermi/Cuntz algebra} [4]. Further applications
of the tilde operators are discussed in Refs. [4-6] and the Appendix.

In this paper  I will focus primarily on the $H$-subalgebra (2.5a) of the
symmetric Fermi/Cuntz algebra,  which is a  single infinite-dimensional
{\it Cuntz algebra}
\setcounter{subsection}{6}
\setcounter{equation}{0}
\begin{equation}
\label{eq2.6a}
b(p) \equiv H(p),\quad b^{\dag}(p) = H(-p),\quad p > 0
\end{equation}
\begin{equation}
\label{eq2.6b}
b(p)b^{\dag}(q) = \delta_{p,q},\quad \sum_{p > 0} b^{\dag}(p)b(p) = 1 - |0\>\<0|
\end{equation}
\vspace{-0.2in}
\begin{equation}
\label{eq2.6c}
b(p)|0\> = \<0|b^{\dag}(p) = 0
\end{equation}
and it should be emphasized that, up to a relabelling, this fermionic Cuntz
algebra is the same Cuntz algebra which is obtained for chiral Bose systems
at large $N$ (see Sec.3). Indeed it is well known [4] that  the Pauli principle
is lost at large $N$ and all systems, Fermi or Bose,  satisfy the same classical
or Boltzmann statistics (with no relations) at large $N$. 

Finally, I will use  standard {\it word notation}  for the reduced basis states 
\setcounter{subsection}{7}
\setcounter{equation}{0}
\begin{equation}
\label{eq2.7a}
b_w \equiv b(p_1)\dots b(p_n),\quad w \equiv p_1\dots p_n
\end{equation}
\vspace{-0.2in}
\begin{equation}
\label{eq2.7b}
[w] = n,\quad \{w\} = \sum_{i=1}^n p_i
\end{equation}
\begin{equation}
\vspace{-0.2in}
\label{eq2.7c}
|w\> = b_w{}^{\dag}|0\> = b^{\dag}(p_n)\dots b^{\dag}(p_1) |0\> = 
H(-p_n)\dots H(-p_1)
|0\>
\end{equation}
\vspace{-0.1in}
\begin{equation}
\label{eq2.7d}
\<w'|w\> = \<0| b_{w'} b_w{}^{\dag} |0\> = \delta_{w',w}
\end{equation}
where $[w]$ and $\{w\}$ are respectively the length and the weight of the word $w$.
Note that, in the present application,  the weight of $w$ is the (mode) level
of the state $|w\>$.  It will also be helpful to recall [4,8] that the reduced basis
states  (2.7) correspond at large $N$ to the unreduced ground state and (dominant)
unreduced adjoint basis states
\renewcommand{\theequation}{\thesection.\arabic{equation}}
\setcounter{equation}{7}
\begin{equation}
\label{eq2.8}
|w.\>_{rs} \equiv \sqrt{N} \left( \frac {b(p_n)^{\dag}}{\sqrt{N}} \dots \frac
{b(p_1)^{\dag}}{\sqrt{N}} \right)_{rs} |0.\>
\end{equation}
whose norm is $O(N^0)$ at large $N$.

\setcounter{subsection}{1}
\subsection{Affine free algebra}
\label{sec2.2}

Let us now return to the unreduced theory, where we may follow Bardakci and
Halpern [9]  to  consider the modes $\{J(m)\}$ of the $\mathfrak{su}(N)$ currents formed
from the unreduced chiral adjoint fermion:
\renewcommand{\theequation}{\thesection.\arabic{subsection}\alph{equation}}
\setcounter{subsection}{9}
\setcounter{equation}{0}
\begin{equation}
\label{eq2.9a}
J_a(m) \equiv \frac {1}{2} \sum_p : H_b(p)(T_a{}^{\mbox{adj}})_{bc}H_c(m-p):
\end{equation}
\vspace{-0.2in}
\begin{equation}
\label{eq2.9b}
(T_a{}^{\mbox{adj}})_{bc} = -if_{abc},\quad J_a(m)^{\dag} = J_a(-m),\quad 
a,b,c = 1\dots N^2
\end{equation}
\begin{equation}
\label{eq2.9c}
J(m)_{rs} \equiv J_a(m)(T_a)_{rs} = -\sum_p : H(p)_{rt}H(m-p)_{ts} :\,= 
J(-m)_{sr}{}^{\dag}
\end{equation}
\vspace{-0.2in}
\begin{equation}
\label{eq2.9d}
: H(p)H(q) :\,\equiv -\theta(p > 0)H(q)H(p) + \theta(p<0)H(p)H(q).
\end{equation}
These currents satisfy the algebra of affine $\mathfrak{su}(N)$ at invariant affine
level $N$, whose matrix form is:
\setcounter{subsection}{10}
\setcounter{equation}{0}
\begin{equation}
\label{eq2.10a}
[J(m)_{rs},J(u)_{uv}] = \delta_{rv}J(m+n)_{us} - \delta_{us}J(m+n)_{rv} +
Nm\delta_{rv}\delta_{us}\delta_{m+n,0}
\end{equation}
\begin{equation}
\label{eq2.10b}
[J(m)_{rs},H(p)_{uv}] = \delta_{rv}H(p+m)_{us} - \delta_{us}H(p+m)_{rv}.
\end{equation}
To find the reduced form of these currents at large $N$, we will need the more
explicit forms:
\setcounter{subsection}{11}
\setcounter{equation}{0}
\begin{equation}
\label{eq2.11a}
J(0)_{rs} = \sum_{p>0} (H(-p)_{ts}H(p)_{rt} - H(-p)_{rt}H(p)_{ts})
\end{equation}
\vspace{-0.2in}
\begin{equation}
\label{eq2.11b}
J(m\ne 0)_{rs} = -\sum_p H_{rt}(p)H_{ts}(m-p).
\end{equation}
These expressions show that the non-zero modes are {\it densities} (matrix products), while
the zero mode involves a term which is a {\it twisted density}.

A key result of Ref. [8] is that twisted densities are {\it subleading} under
large $N$ reduction, and so we obtain the explicit form of the modes of
the {\it reduced currents}
\setcounter{subsection}{12}
\setcounter{equation}{0}
\begin{equation}
\label{eq2.12a}
\frac {1}{N} J(m)_{rs} \rightarrow J(m)
\end{equation}
\begin{equation}
\vspace{-0.1in}
\label{eq2.12b}
J(0) = -\sum_{p>0} H(-p)H(p) = |0\>\<0| - 1,\quad J(0)^2 = -J(0)
\end{equation}
\vspace{-0.2in}
\begin{equation}
\label{eq2.12c}
J(m \ne 0) = -\sum_p H(p)H(m-p)
\end{equation}
in terms of the reduced fermion modes.  The second form of $J(0)$ in
(2.12b) follows from (2.5b). Tilde partners of the reduced current modes 
can also be obtained, but I will not study these here.

The reduced current modes (2.12) can be written in the more useful form:
\setcounter{subsection}{13}
\setcounter{equation}{0}
\begin{equation}
\label{eq2.13a}
J(m \ge 0) = -\sum_{p > 0} H(m-p)H(p)
\end{equation}
\vspace{-0.2in}
\begin{equation}
\label{eq2.13b}
J(m \le 0) = -\sum_{p>0} H(-p)H(m+p)
\end{equation}
\vspace{-0.1in}
\begin{equation}
\label{eq2.13c}
J(m)^{\dag} = J(-m),\quad J(m \ge 0) |0\> = \<0| J(m\le 0) = 0
\end{equation}
\begin{equation}
\label{eq2.13d}
J(0) |\alpha\> = -|\alpha\>,\quad \<\alpha| J(0) = -\<\alpha|,\quad \forall\ 
|\alpha\> \ne |0\>.
\end{equation}
To obtain (2.13a,b) from (2.12c), use the fermionic Cuntz algebra (2.5a)
to verify the identities
\setcounter{subsection}{14}
\setcounter{equation}{0}
\begin{equation}
\label{eq2.14a}
\sum_{p>m} H(p)H(m-p) = 0\, \mbox{ for } m > 0
\end{equation}
\vspace{-0.2in}
\begin{equation}
\label{eq2.14b}
\sum_{p>0} H(p)H(m-p) = 0\, \mbox{ for } m < 0
\end{equation}
and use the second form of J(0) in (2.12b) to verify (2.13d). 

Because the reduced currents (2.13) satisfy only free-algebraic relations,
I will sometimes refer to them as {\it free currents}.  For example, the free algebra
of the free currents with the reduced fermions
\renewcommand{\theequation}{\thesection.\arabic{equation}}
\setcounter{equation}{14}
\begin{equation}
\label{eq2.15}
J(m \ge 0)H(p<0) = H(p>0)J(m\le 0) = -H(p+m)
\end{equation}
follows from (2.13) and the fermionic Cuntz algebra.

Moreover, we may use (2.13) and (2.15) to obtain the free algebra of the
free current modes with themselves, which we call {\it affine free algebra}
at reduced level $\hat{x}=1$:
\begin{equation}
\label{eq2.16}
J(m \ge 0)J(n \le 0) = -J(m+n) + m\delta_{m+n,0},\quad m,n \in {\mathbb Z}.
\end{equation}
The affine free algebra (2.16), which describes the large $N$ limit of the
$\mathfrak{su}(N)$ affine Lie algebra (2.10a) in the adjoint sector, is one of the central
results of this paper.  It should be emphasized that this  result is an
example in the class of  so-called {\it interacting Cuntz algebras}, which appear
quite generally in interacting Bose systems [4-6] at large $N$.

To obtain a realization of affine free algebra at higher integer level, begin
with $\hat{x}\in{\mathbb Z}_+$  chiral fermions in the adjoint.  For brevity, I give only
the reduced results for this case:
\renewcommand{\theequation}{\thesection.\arabic{subsection}\alph{equation}}
\setcounter{subsection}{17}
\setcounter{equation}{0}
\begin{equation}
\label{eq2.17a}
H_i(p>0)H_j(q<0) = \delta_{ij}\delta_{p+q,0},\quad i,j = 1\dots {\hat x}
\end{equation}
\begin{equation}
\label{eq2.17b}
\sum_{i=1}^{\hat x} \sum_{p>0} H_i(-p)H_i(p) = 1 - |0\>\<0|
\end{equation}
\vspace{-0.3in}
\begin{equation}
\label{eq2.17c}
J(m) \equiv \sum_{i=1}^{\hat x} J_i(m)
\end{equation}
\vspace{-0.2in}
\begin{equation}
\label{eq2.17d}
J_i(m \ge 0) \equiv -\sum_{p>0} H_i(m-p)H_i(p),\ J_i(m \le 0) \equiv 
-\sum_{p>0} H_i(-p)H_i(m+p)
\end{equation}
\vspace{-0.3in}
\begin{equation}
\label{eq2.17e}
J_i(m)^{\dag} = J_i(-m),\quad J(m)^{\dag} = J(-m),\quad J(0) = |0\>\<0| - 1
\end{equation}
\begin{equation}
\label{eq2.17f}
J(m\ge 0)|0\> = J_i(m \ge 0)|0\> = 0,\ \<0| J(m \le 0) = \<0|J_i(m \le 0) = 0
\end{equation}
\begin{equation}
\label{eq2.17g}
J_i(m \ge 0) H_j(p<0) = H_i(p>0)J_j(m \le 0) = -\delta_{ij}H_i(p+m)
\end{equation}
\begin{equation}
\label{eq2.17h}
J(m \ge 0)H_i(p>0) = H_i(p>0)J(m\le 0) = -H_i(p+m)
\end{equation}
\begin{equation}
\label{eq2.17i}
J(m \ge 0)J(n \le 0) = -J(m+n) + {\hat x}m\delta_{m+n,0}.
\end{equation}
At reduced level $\hat{x}$, the affine free algebra (2.17i) is the result quoted
in the Introduction.

The representation theory of affine free algebra, including the possibility
of non-integer reduced level $\hat{x}$, is discussed in the following subsections.

\setcounter{subsection}{2}
\subsection{Free affine primary states and modules}
\label{sec2.3}

Following the usual convention, I will define a {\it free affine primary state}
$|FAP\>$ of the affine free algebra (2.17i) at arbitrary reduced level $\hat{x}$
as any state which satisfies
\renewcommand{\theequation}{\thesection.\arabic{equation}}
\setcounter{equation}{17}
\begin{equation}
\label{eq2.18}
J(m \ge 0)|FAP\> = -\delta_{m,0}t|FAP\>,\quad t = 0 \mbox{ or } 1
\end{equation}
and I will assume in what follows that each free affine primary state
is normalized to 1. Only the indicated values of the eigenvalue $t$  are
allowed because  $-J(0) = J(0)^2$  is  a projection operator . We may always
introduce a {\it free affine ground state} $| 0\>$ which is a free affine primary state
with $t = 0$ . For the fermionic realizations  at positive integer $\hat{x}$ we
find with (2.13c,d) that
\begin{equation}
\label{eq2.19}
t = \begin{cases}
0 &\mbox{for $|0\>$} \\
1 &\mbox{for all others}
\end{cases}
\end{equation}
so that, in these realizations,  the fermion ground state $|0\>$ is  the only
free affine primary state with $t =0$. 

The {\it free affine module} of any free affine primary state is constructed as
usual with the negative modes of the reduced currents, and the affine free
algebra tells us that all the free affine secondaries of each module have $t=1$
\renewcommand{\theequation}{\thesection.\arabic{subsection}\alph{equation}}
\setcounter{subsection}{20}
\setcounter{equation}{0}
\begin{equation}
\label{eq2.20a}
(J(0) + 1)J(-m) = 0,\ m \ge 0
\end{equation}
\begin{equation}
\label{eq2.20b}
(J(0)+1)J(-m_1)\dots J(-m_n)|FAP\> = 0,\ n \ge 1
\end{equation}
independent of the free affine primary state. 

Using (2.20) and the affine free algebra (2.17i) at arbitrary reduced
level $\hat{x}$, it is straightforward to compute the norms of all the states
in  each free affine module
\renewcommand{\theequation}{\thesection.\arabic{equation}}
\setcounter{equation}{20}
\begin{equation}
\label{eq2.21}
\|J(-m_1)\dots J(-m_n)|FAP\>\|^2 = (t+m_n{\hat x}) \prod_{i=1}^n 
(1+m_i{\hat x}) > 0
\mbox{ for } {\hat x} > 0.
\end{equation}
This surprisingly simple result shows that  all the norms are positive
and there are no null states in the free affine modules so long as the
level $\hat{x}$ is strictly greater than zero. Although I will not pursue this
subject here, the result (2.21) suggests that unitary realizations of the affine
free algebra exist for all $\hat{x}> 0$.

As concrete examples, I list here the lowest (normalized) free affine primary
states at reduced level $\hat{x}=1$
\renewcommand{\theequation}{\thesection.\arabic{subsection}\alph{equation}}
\setcounter{subsection}{22}
\setcounter{equation}{0}
\begin{equation}
\label{eq2.22a}
|\Delta=0,0\> = |0\>,\quad |\Delta = 1/2,1\> = H(-1/2)|0\>
\end{equation}
\begin{equation}
\label{eq2.22b}
|\Delta = 3/2,1\> = \frac {1}{\sqrt{2}} (H(-1/2)^3 - H(-3/2))|0\>
\end{equation}
and the lowest free affine primary states at reduced level $\hat{x}\in{\mathbb Z}_+$
\setcounter{subsection}{23}
\setcounter{equation}{0}
\begin{equation}
\label{eq2.23a}
|\Delta=0,0\> = |0\>,\quad |\Delta_i = 1/2,1\> = H_i(-1/2)|0\>,\quad i=1\dots {\hat x}
\end{equation}
\begin{equation}
\label{eq2.23b}
|\Delta_{i \ne j}=1,1\> = H_i(-1/2)H_j(-1/2)|0\>,\quad 1 \le i \ne j \le {\hat x}
\end{equation}
\begin{equation}
\label{eq2.23c}
|\Delta_{ii} = 1,1\> = \frac {1}{\sqrt{2}} (H_i(-1/2)^2 - 
H_{i+1}(-1/2)^2)|0\>,\quad i =
1\dots {\hat x}-1.
\end{equation}
The labelling of these states will be explained in the following subsections,
where we will also study the partition functions and characters  associated
to positive integer level of the free affine algebra.

As a final topic in this subsection, I want to distinguish between two types
of free affine primary states which I will call ``ordinary" and 
``extraordinary". The
distinction is based on the fact that, as we shall see, we are finding more
free affine primary states than there are adjoint affine primary states in the
unreduced theory
\renewcommand{\theequation}{\thesection.\arabic{equation}}
\setcounter{equation}{23}
\begin{equation}
\label{eq2.24}
\left( \frac {J(m>0)}{N} \right)_{rt} C(p_1 ,\ldots ,p_n ) \left\{ \sqrt{N} \frac {H(-p_1)}{\sqrt{N}} 
\dots \frac{H(-p_n)}{\sqrt{N}} \right\}_{ts} |0.\> = 0
\end{equation}
at any finite value of $N$.   Examples of adjoint affine primary states
at finite $N$ are easily found in the case of one adjoint fermion:
\renewcommand{\theequation}{\thesection.\arabic{subsection}\alph{equation}}
\setcounter{subsection}{25}
\setcounter{equation}{0}
\begin{equation}
\label{eq2.25a}
\left( \frac {J(m>0)}{N}\right)_{rs} |0.\> = \left( \frac {J(m>0)}{N} 
H(-1/2)\right)_{rs}
|0.\> = 0
\end{equation}
\vspace{-0.1in}
\begin{equation}
\label{eq2.25b}
\left( \frac {J(m\ge 2)}{N} \sqrt{N} \left\{ \left( \frac 
{H(-1/2)}{\sqrt{N}} \right)^3 -
H(-3/2)\right\}\right)_{rs} |0.\> = 0.
\end{equation}
Then the free affine primary states in (2.22a) are {\it ordinary}
in the sense that they are in one-to-one correspondence ($H\leftrightarrow H_{rs}$)
with these adjoint affine
primary states at all finite values of $N$.  Similarly, one finds that the adjoint
states which correspond to the free affine primary states
in Eq. (2.23) are in fact affine primary states at finite $N$, so these free
affine primary states are also ordinary.  

On the other hand, we find after some algebra that the unreduced adjoint state
which corresponds to the free affine primary state (2.22b) satisfies
\renewcommand{\theequation}{\thesection.\arabic{subsection}\alph{equation}}
\setcounter{subsection}{26}
\setcounter{equation}{0}
\begin{equation}
\label{eq2.26a}
\left( \frac {J(m\ge 2)}{N} \sqrt{N} \left\{ \left( \frac 
{H(-1/2)}{\sqrt{N}} \right)^3 -
H(-3/2)\right\}\right)_{rs} |0.\> = 0
\end{equation}
\vspace{-0.1in}
\begin{eqnarray}
\label{eq2.26b}
& &\left( \frac {J(1)}{N} \sqrt{N} \left\{ \left( \frac 
{H(-1/2)}{\sqrt{N}} \right)^3 -
H(-3/2)\right\}\right)_{rs} |0.\> \nonumber \\
& &\hskip 0.5 in = \left\{ \frac {1}{N^2} H(-1/2)_{rs} - \frac
{\delta_{rs}}{\sqrt{N}} \mbox{Tr} \left( \frac {H(-1/2)}{\sqrt{N}} 
\right)\right\} |0.\>
\end{eqnarray}
so that this unreduced state is not an affine primary state at any finite $N$.
The trace term in (2.26) is an artifact of our inclusion  of the extra $\mathfrak{u}(1)$
fermion, but the first term would appear even for a traceless fermion. Of course
the norm of this state goes to zero smoothly at large $N$, in fact  as $O(N^{-1})$,
so it is no surprise that the reduced state (2.22b) is a free affine primary 
state -- albeit an {\it extraordinary} one which does not correspond to an adjoint
affine primary state at any finite value of $N$.

The lesson here is that new extraordinary free affine primary states can and do occur
in the large $N$ limit (accompanied in fact by new null states such as the limit
of (2.26)), and these extraordinary  states are included on an equal footing
with the ordinary free affine primary states in our free-algebraic formulation.

\setcounter{subsection}{3}
\subsection{Reduced fermionic $\mathfrak{sl}(2)$}
\label{sec2.4}

I turn next to study the fate of the Virasoro algebra in large $N$ matrix mechanics. 

Before considering any particular  theory,  some general remarks will be helpful.
In the first place,  the Virasoro generators of  general  matrix conformal field
theories  are $\mathfrak{su}(N)$-invariant operators  which comprise particular  examples of
so-called {\it trace class operators} in each theory.   It is known [4] that  algebraic
relations among trace class operators  and  algebraic relations  of  trace class
operators with the densities of the theory are preserved under large $N$
reduction-- but the composite structure of reduced trace class operators is
not obtained directly in the reduction procedure. This is called the  opacity
problem in Ref. [4]. To obtain the composite structure of a given reduced trace class
operator, it is necessary to  solve [4,8] the algebraic relations  of that operator
with the reduced fundamental densities of the theory. 

As a concrete example  I will consider  first the theory of a single adjoint
fermion, which has conformal weight 1/2 under the Virasoro algebra.  It follows
that, if the reduced Virasoro generators $\{L_F(m)\}$ are to exist at all, they must
solve the reduced commutation relations
\renewcommand{\theequation}{\thesection.\arabic{equation}}
\setcounter{equation}{26}
\begin{equation}
\label{eq2.27}
[L_F(m),H(p)] = -(\frac{m}{2} + p)H(p+m),\quad \forall\ p \in {\mathbb Z} + 1/2
\end{equation}
which have the same form as the corresponding unreduced relations.

It  is clear  on general grounds however that the Virasoro generators and
hence their reduced counterparts cannot be well-defined for  $|m|>1$, so that
something will prevent us from solving (2.27) beyond  the $\mathfrak{sl}(2)$ subalgebra.
The reason is that the central charge of the unreduced theory is $c =N^2/2$,
and more generally the central charge  of any matrix conformal field theory
grows as $O(N^2)$ at large $N$. This means that the  Virasoro generators  with
$m\le -2$  would  create states of infinite norm at
large $N$--and this is of course quite intolerable in a rigorous formulation such
as large $N$ matrix mechanics. 

In fact, it is easy to find the  implied  inconsistency in our example.  Use
the reduced commutation relations (2.27) to compute the commutator of $L_F(m)$
with the fermion Cuntz algebra (2.5a):
\renewcommand{\theequation}{\thesection.\arabic{subsection}\alph{equation}}
\setcounter{subsection}{28}
\setcounter{equation}{0}
\begin{eqnarray}
\label{eq2.28}
0 &=& [L_F(m),H(p>0)H(q<0)] \\
&=& (m/2+p)H(p+m)H(q) + (\frac{m}{2} + q)H(p)H(q+m).
\end{eqnarray}
These relations are consistent with the Cuntz algebra when $|m|\le 1$. For example 
one finds
\renewcommand{\theequation}{\thesection.\arabic{equation}}
\setcounter{equation}{28}
\begin{equation}
\label{eq2.29}
0 = (p+q)H(p)H(q) = (p+q)\delta_{p+q,0}
\end{equation}
when $m=0$.  But the relations (2.28) are inconsistent with Cuntz for
$|m|>1$ -- for example, we find the inconsistent relation
\begin{equation}
\label{eq2.30}
0 = \frac {1}{2} H(p)H(3/2),\quad p \in {\mathbb Z} + \frac {1}{2}
\end{equation}
when $m= 2$ and $q = -1/2$.

So we must be satisfied at large $N$ to construct reduced generators only for
the $\mathfrak{sl}(2)$ subalgebra of the original Virasoro algebra.  To proceed in this example,
it is helpful to recall [4] the following lemma
\renewcommand{\theequation}{\thesection.\arabic{subsection}\alph{equation}}
\setcounter{subsection}{31}
\setcounter{equation}{0}
\begin{equation}
\label{eq2.31a}
A(B) \equiv \sum_w b_w{}^{\dag} B b_w = B + \sum_{p>0} b^{\dag}(p)Bb(p) + \dots
\end{equation}
\vspace{-0.2in}
\begin{equation}
\label{eq2.31b}
[A(B),b(p)] = -b(p)B,\quad [A(B),b^{\dag}(p)] = Bb^{\dag}(p)
\end{equation}
where $B$, called the {\it kernel} of $A$,  is an arbitrary function of the Cuntz operators
$b,b^{\dag}$. The operation $\sum_w b_w{}^{\dag}(...)b_w$ in (2.31a) is
called the {\it dressing} of the kernel $B$.
Then one finds the explicit form of the reduced $sl(2)$ generators
\setcounter{subsection}{32}
\setcounter{equation}{0}
\begin{equation}
\label{eq2.32a}
L_F(m) = \sum_w b_w{}^{\dag}B_F(m)b_w,\quad |m| \le 1
\end{equation}
\vspace{-0.1in}
\begin{equation}
\label{eq2.32b}
B_F(m) = \sum_{p>0} (\frac{m}{2} + p)H(-p)H(p+m)
\end{equation}
as the solution of (2.27 ) for $|m|\le 1$.

Using the fermionic Cuntz algebra (2.5a), we also verify the following properties
of the kernels
\setcounter{subsection}{33}
\setcounter{equation}{0}
\begin{equation}
\label{eq2.33a}
B_F(m)^{\dag} = B_F(-m),\quad B_F(m)|0\> = \<0| B_F(m) = 0
\end{equation}
\begin{equation}
\label{eq2.33b}
H(p>0)B_F(m) = -B_F(m)H(p<0) = (\frac{m}{2} +p)H(p+m)
\end{equation}
\begin{equation}
\label{eq2.33c}
[B_F(m),B_F(n)] = (m-n)B_F(m+n)
\end{equation}
\begin{equation}
\label{eq2.33d}
[L_F(m),B_F(n)] = (m-n)B_F(m+n)
\end{equation}
and these properties allow us to check that our reduced operators generate
the expected  $\mathfrak{sl}(2)$ 
\setcounter{subsection}{34}
\setcounter{equation}{0}
\begin{equation}
\label{eq2.34a}
[L_F(m),L_F(n)] = \sum_w b_w{}^{\dag} [L_F(m),B_F(n)]b_w
\end{equation}
\begin{equation}
\label{eq2.34b}
= (m+n)L_F(m+n),\quad |m|,\ |n| \le 1
\end{equation}
\begin{equation}
\label{eq2.34c}
L_F(m)^{\dag} = L_F(-m),\quad L_F(m)|0\> = \<0| L_F(m)=0.
\end{equation}
Note that the reduced  fermionic ground state is an $\mathfrak{sl}(2)$-invariant state, as
in the unreduced theory, so that we may derive $\mathfrak{sl}(2)$ Ward identities (see
the Appendix) for the reduced ground state averages -- which are the large $N$ limit
of the unreduced traces. 

Moreover, we may use (2.27) for $|m|\!\leq\!1$ and the explicit form (2.13) of the free current
modes  to show that the free currents are $(1,0)$ operators under the reduced
$\mathfrak{sl}(2)$
\renewcommand{\theequation}{\thesection.\arabic{equation}}
\setcounter{equation}{34}
\begin{equation}
\label{eq2.35}
[L_F(m),J(n)] = -nJ(m+n)
\end{equation}
as they are in the unreduced theory.   The action of $L_F(0)$ on the reduced
basis states (2.7c)  is  also not surprising 
\renewcommand{\theequation}{\thesection.\arabic{subsection}\alph{equation}}
\setcounter{subsection}{36}
\setcounter{equation}{0}
\begin{equation}
\label{eq2.36a}
[L_F(0),b^{\dag}(p)] = pb^{\dag}(p)
\end{equation}
\begin{equation}
\label{eq2.36b}
(L_F(0)-\Delta(w))|w\> = \<w| (L_F(0) - \Delta(w)) = 0,\quad \Delta(w) = \{w\}
\end{equation}
where $\{w\}$ in (2.7b) is the weight (sum of the mode numbers) of the word $w$.
Since $L_F(0)$
commutes with $J(0)$, we may then label the free affine primary states as follows:
\setcounter{subsection}{37}
\setcounter{equation}{0}
\begin{equation}
\label{eq2.37a}
J(m \ge 0)|\Delta,t\> = |\Delta,t\> \delta_{m,0}(-t)
\end{equation}
\begin{equation}
\label{eq2.37b}
L_F(0) |\Delta,t\> = |\Delta,t\>\Delta.
\end{equation}
This is in fact the labelling employed in the examples (2.22 ) and (2.23).
Of course, the eigenvalues $\Delta$ are real because $L_F(0)$ is hermitian,
and, similarly, Schur's lemma guarantees that we may take all distinct free
affine primary states (and their modules) to be orthogonal. 

It is instructive to consider the action of the reduced $\mathfrak{sl}(2)$ generator
$L_F(1)$ on a
free affine primary state. Using  (2.35-37) and the reduced $\mathfrak{sl}(2)$ 
algebra, it is not
difficult to see that
\setcounter{subsection}{38}
\setcounter{equation}{0}
\begin{equation}
\label{eq2.38a}
J(m\ge 0)(L_F(1)|\Delta,t\>) = (L_F(1)|\Delta,t\>) \delta_{m,0}(-t)
\end{equation}
\begin{equation}
\label{eq2.38b}
L_F(0)(L_F(1)|\Delta,t\>) = (L_F(1)|\Delta,t\>)(\Delta-1).
\end{equation}
These relations tell us either that a) the state $L_F(1)|\Delta, t\>$ vanishes, in
which case the free affine primary state $| \Delta, t\>$ was an $\mathfrak{sl}(2)$ primary state,
or b) the state $L_F(1)|\Delta,t\>$  is another (lower) free affine primary state,
and the original free affine primary state $|\Delta, t\>$  was not $\mathfrak{sl}(2)$ primary.
In fact, both cases can and do occur, as seen for the examples at 
$\hat{x}=1$ in Eq.(2.22):
\setcounter{subsection}{39}
\setcounter{equation}{0}
\begin{equation}
\label{eq2.39a}
L_F(1)|\Delta=t=0\> = L_F(1)|\Delta = 1/2,t = 1\> = 0
\end{equation}
\begin{equation}
\label{eq2.39b}
L_F(1)|\Delta=3/2,t=1\> = -\frac {1}{\sqrt{2}} |\Delta=1/2,t=1\>.
\end{equation}
Of course, any free affine primary state which is not $\mathfrak{sl}(2)$ primary must be one
of the extraordinary free affine primary states (see Subs. 2.3) -- which does
not correspond to an unreduced affine primary state at any finite value of $N$.
The reason  is that, up to the extra $\mathfrak{u}(1)$ fermion, the unreduced free-fermion
Virasoro generators are equal to those of the affine-Sugawara construction
[9,12,15-17] on the unreduced fermionic currents  (2.9) at level $N$ of $\mathfrak{su}(N)$
-- and for the affine-Sugawara construction all affine primary states are also
Virasoro primary. 

The generalization of the reduced $\mathfrak{sl}(2)$ generators to the case of an arbitrary
number $\hat{x}$ of adjoint fermions is 
\setcounter{subsection}{40}
\setcounter{equation}{0}
\begin{equation}
\label{eq2.40a}
L_F(m) = \sum_w b_w{}^{\dag} B_F(m)b_w,\quad |m| \le 1
\end{equation}
\vspace{-0.2in}
\begin{equation}
\label{eq2.40b}
B_F(m) = \sum_{i=1}^{\hat x} \sum_{p>0} (\frac{m}{2} + p)H_i(-p)H_i(p+m)
\end{equation}
\vspace{-0.1in}
\begin{equation}
\label{eq2.40c}
b_w = H_{i_1}(p_1)\dots H_{i_n}(p_n),\quad w = (i_1p_1,\dots,i_np_n)
\end{equation}
\begin{equation}
\label{eq2.40d}
[w] = n,\quad \{w\} = \sum_{i=1}^n p_i
\end{equation}
and, except for an extra label ${i=1...\hat{x}}$ on the $b^{\dag}$'s in (2.36a), there is no change
in any of the  results and conclusions above. In particular,  we know that the
ordinary free affine primary states in (2.23 ) are primary states under this 
$\mathfrak{sl}(2)$.

\setcounter{subsection}{4}
\subsection{Free-algebraic partition functions and characters}
\label{sec2.5}

Starting with $\hat{x}\in{\mathbb Z_+}$  reduced (Cuntz-algebraic) chiral
fermions and the corresponding reduced $\mathfrak{sl}(2)$ generator $L_F(0)$ 
in (2.40), one can define and evaluate the 
{\it reduced fermionic partition function}
\setcounter{subsection}{41}
\setcounter{equation}{0}
\begin{eqnarray}
\label{eq2.41a}
Z_F(z) &\equiv& \mbox{Tr}(z^{L_F(0)}) = \sum_w z^{\{w\}} = \frac 
{1}{1-{\hat x} \sum_{q>0}
z^q}\\
&=& \frac {1-u^2}{1-{\hat x}u-u^2} = \sum_{n=0}^{\infty} f_n({\hat 
x})u^n,\quad z = u^2
\end{eqnarray}
where the positive numbers $f_n(\hat{x})$ (see below) count the reduced fermionic
states at  mode level $L_F(0) = n/2$.   The radius of convergence $u_0(\hat{x})$ of
this  power series and the corresponding 2-d {\it limiting} (Hagedorn) 
{\it temperature}
$\beta_0(\hat{x})^{-1}$ of this system are
\setcounter{subsection}{42}
\setcounter{equation}{0}
\begin{equation}
\label{eq2.42a}
z = e^{-\beta},\quad u = e^{-\beta/2}
\end{equation}
\begin{equation}
\label{eq2.42b}
u_0({\hat x}) = e^{-\beta_0({\hat x})/2} = \frac {1}{2} (\sqrt{{\hat 
x}^2+4} - {\hat x})
< 1
\end{equation}
\begin{equation}
\label{eq2.42c}
e^{\beta_0({\hat x})/2} = \frac {1}{2} (\sqrt{{\hat x}^2+4} + {\hat x})
\end{equation}
and the ratio in (2.41b) provides an analytic continuation to temperatures beyond
the limiting temperature. 

The radius of convergence in (2.42) is less than the usual radius of convergence
(at $u =z =1$) of conventional string partition functions (say  for any finite number
of abelian chiral world-sheet fermions).  This tells us that we are dealing here
with a great many more states than are present in the conventional string partition
functions, and it is this fact that gives us a finite 2-d limiting temperature --
instead of the infinite 2-d limiting temperature obtained in the conventional case.
Relative to the counting of states for  conventional chiral world-sheet fermions,
the large number of reduced fermionic states in the  reduced large N theory is due
to the classical or Boltzmann statistics  of the reduced (Cuntz-algebraic) fermions,
that is, the loss of the Pauli principle at large $N$. ( The large number of states
is not associated to the $N^2$ degrees of freedom in the $r,s$ labels of the unreduced
matrix fermions, since these degrees of freedom are modded out in the reduction
 procedure.)

To evaluate the integers $\{f_n(\hat{x})\}$ in (2.41b), consider first the 
{\it generalized Fibonacci numbers} $F_n(\hat{x})$ 
\setcounter{subsection}{43}
\setcounter{equation}{0}
\begin{equation}
\label{eq2.43a}
(1-{\hat x}u-u^2)^{-1} = \sum_{n=0}^{\infty} F_n({\hat x})u^n
\end{equation}
\begin{equation}
\label{eq2.43b}
F_0({\hat x}) = 1,\quad F_1({\hat x}) = {\hat x},\quad  F_{n+2}({\hat x}) = 
{\hat x}F_{n+1}({\hat
x}) + F_n({\hat x}),\quad n \ge 0
\end{equation}
\begin{equation}
\label{eq2.43c}
F_n({\hat x}) = \frac {1}{\sqrt{{\hat x}^2+4}} \left( \left(\frac 
{{\hat x}+\sqrt{{\hat
x}^2+4}}{2} \right)^{n+1} - \left( \frac {{\hat x}-\sqrt{{\hat x}^2+4}}{2}
\right)^{n+1}\right)
\end{equation}
\vspace{-0.1in}
\begin{equation}
\label{eq2.43d}
F_n (\hat{x}){}_{\stackrel{\mbox{\Large $\sim$}}{n \gg 1}}\ e^{n\beta_0({\hat x})/2} = e^{n \ln\left( \frac
{{\hat x}+\sqrt{{\hat x}^2+4}}{2}\right)}
\end{equation}
which are defined so that $F_n(1)$ is  the nth Fibonacci number. Then one finds
from (2.41b) that 
\setcounter{subsection}{44}
\setcounter{equation}{0}
\begin{equation}
\label{eq2.44a}
f_0({\hat x}) = 1,\quad f_{n+1}({\hat x}) = {\hat x}F_n({\hat x}),\quad n \ge 0
\end{equation}
\begin{equation}
\label{eq2.44b}
f_{n+1}(1) = F_n(1),\quad f_n({\hat x})\ {}_{\stackrel{\mbox{\Large $\sim$}}{n \gg 1}}\ 
e^{n\beta_0({\hat
x})/2}
\end{equation}
and this  $O(\exp(an))$ growth at large $n$ should be compared with the familiar
$O(\exp(b\sqrt{n}))$ growth found in conventional string partition functions.  The
same qualitative behavior is found (see Sec.3) in the partition functions of
reduced chiral bosons.

I turn next to the {\it free affine character}
\renewcommand{\theequation}{\thesection.\arabic{equation}}
\setcounter{equation}{44}
\begin{equation}
\label{eq2.45}
\chi_{_{\Delta= \frac {n}{2}}}(z) \equiv \mbox{Tr}_{\Delta=\frac 
{n}{2}} (z^{L_F(0)}) =
\frac {u^n(1-u^2)}{1-2u^2}
\end{equation}
which sums over the states of the module  of any free affine primary state at
mode level $\Delta = n/2$. One may expand the  reduced fermionic partition functions
(2.41) in terms of the free affine characters
\begin{equation}
\label{eq2.46}
Z_F(z) = \sum_{n=0}^{\infty} a_n({\hat x}) \chi_{_{\frac {n}{2}}}(z),\quad \frac 
{1-2u^2}{1-{\hat
x}u-u^2} = \sum_{n=0}^{\infty} a_n({\hat x})u^n
\end{equation}
so that the integers $\{a_n(\hat{x})\}$ count the number of free affine primary states at
mode level $n/2$ in the reduced fermionic Hilbert space. From the last part of
(2.46) one finds that
\renewcommand{\theequation}{\thesection.\arabic{subsection}\alph{equation}}
\setcounter{subsection}{47}
\setcounter{equation}{0}
\begin{equation}
\label{eq2.47a}
a_0({\hat x}) = 1,\quad a_1({\hat x}) = {\hat x},\quad a_2({\hat x}) = {\hat x}^2-1
\end{equation}
\begin{equation}
\vspace{-0.2in}
\label{eq2.47b}
a_{n+2}({\hat x}) = ({\hat x}^2-1)F_n({\hat x}) + {\hat 
x}F_{n-1}({\hat x}) = {\hat
x}a_{n+1}({\hat x}) + a_n({\hat x}),\quad n \ge 1
\end{equation}
\vspace{-0.1in}
\begin{equation}
\label{eq2.47c}
a_{n+2}(1) = F_{n-1}(1),\quad a_n({\hat x})\ {}_{\stackrel{\mbox{\Large $\sim$}}{n \gg 
1}}\ F_n({\hat x})
\end{equation}
and these numbers are in agreement with the low-lying examples in
Eqs. (2.22) and (2.23).

Further discussion of affine free algebra is found in Sec.4.

\section{Chiral Adjoint Bosons}
\label{sec3}

In this section I will develop the corresponding large $N$ theory of chiral 
adjoint bosons which, up to a point and except for some subtlety with the zero
modes, is simpler than that given above for the large $N$ fermions.

\subsection{Cuntz algebra and reduced bosonic partition functions}
\label{sec3.1}

I will begin with the unreduced theory of a single chiral adjoint boson in sector $k$   
\setcounter{subsection}{1}
\setcounter{equation}{0}
\begin{equation}
\label{eq3.1a}
[\pi_a(m),\pi_b(n)] = \delta_{ab}m\delta_{m+n,0},\quad a,b = 1\dots N^2
\end{equation}
\vspace{-0.2in}
\begin{equation}
\label{eq3.1b}
\pi_a(m)^{\dag} = \pi_a(-m),\quad m \in {\mathbb Z}
\end{equation}
\vspace{-0.2in}
\begin{equation}
\label{eq3.1c}
(\pi_a(m \ge 0) - \delta_{m,0}k_a)|k.\> = \<.k|(\pi_a(m\le 0) - 
\delta_{m,0}k_a) = 0
\end{equation}
where $\{k_a\}$ are the eigenvalues of the zero modes $\{\pi_a(0)\}$. The matrix forms
of these operators are defined as usual, and we find:
\setcounter{subsection}{2}
\setcounter{equation}{0}
\begin{equation}
\label{eq3.2a}
\pi(m)_{rs} = \pi_a(m)(T_a)_{rs},\quad k_{rs} = k_a(T_a)_{rs},\quad r,s = 1\dots N
\end{equation}
\vspace{-0.1in}
\begin{equation}
\label{eq3.2b}
[\pi(m)_{rs},\pi(n)_{uv}] = \delta_{rv}\delta_{su}m\delta_{m+n,0}
\end{equation}
\vspace{-0.1in}
\begin{equation}
\label{eq3.2c}
[\pi(m)_{rs},k_{uv}] = 0,\quad \pi(m)_{rs}{}^{\dag} = \pi(-m)_{sr}
\end{equation}
\vspace{-0.1in}
\begin{equation}
\label{eq3.2d}
\{\pi(m\ge 0) - \delta_{m,0}k\}_{rs} |k.\> = \<.k| \{\pi(m\le 0) - 
\delta_{m,0}k\}_{rs} = 0.
\end{equation}
The large $N$ reduction of this system  follows closely the development in Refs.[4,8].

In particular, Eq.(2.29) of Ref.[4] gives us the quasicanonical algebra of the
reduced operators in {\it reduced sector} $k$:
\setcounter{subsection}{3}
\setcounter{equation}{0}
\begin{equation}
\label{eq3.3a}
\frac {\pi(m)_{rs}}{\sqrt{N}} \rightarrow \pi(m),{\tilde \pi}(m),\quad \frac
{k_{rs}}{\sqrt{N}} \rightarrow {\tilde k} = k
\end{equation}
\vspace{-0.2in}
\begin{equation}
\label{eq3.3b}
[\pi(m),{\tilde \pi}(n)] = m \delta_{m+n,0} |k\>\<k|, \quad m,n\in{\mathbb Z}
\end{equation}
\begin{equation}
\vspace{-0.1in}
\label{eq3.3c}
\pi(m)^{\dag} = \pi(-m),\quad {\tilde \pi}(m)^{\dag} = {\tilde \pi}(-m)
\end{equation}
\vspace{-0.1in}
\begin{equation}
\label{eq3.3d}
{\tilde \pi}(0) = \pi(0) ,\quad [\pi(m\neq 0), \pi(0)] = [{\tilde 
\pi}(m\neq 0),\pi(0)] =0
\end{equation}
\vspace{-0.1in}
\begin{equation}
\label{eq3.3e}
(\pi(m\ge 0)-\delta_{m,0}k)|k\> = ({\tilde \pi}(m \ge 0) - 
\delta_{m,0}k)|k\> = 0
\end{equation}
\vspace{-0.1in}
\begin{equation}
\label{eq3.3f}
\<k|(\pi(m\le 0) - k\delta_{m,0}) = \<k|{\tilde \pi}(m \le 0) - 
\delta_{m,0}k) = 0
\end{equation}
\vspace{-0.1in}
\begin{equation}
\label{eq3.3g}
({\tilde \pi}(m \le 0) - \pi(m\le 0))|k\> = \<k|({\tilde \pi}(m\ge 0) 
- \pi(m\ge 0)) = 0.
\end{equation}
Here I have chosen to scale the unreduced eigenvalues so that the reduced
eigenvalues $k$ and orthonormal reduced eigenstates $|k\>$ are finite in the
large $N$ limit. I will also assume that the reduced momenta are defined on
some lattice, so that the reduced eigenstates can be taken orthonormal. Again,
a basis for sector $k$ can be formed by the action on $|k\>$ of the negative modes
of either the untilded operators or the tilded operators. Finally (see Eq.(3.7)
of Ref.[4])  operation on either basis with both sets of positive modes  shows
that the reduced bosonic modes also satisfy the free-algebraic relations
\setcounter{subsection}{4}
\setcounter{equation}{0}
\begin{equation}
\label{eq3.4a}
\pi(m>0)\pi(n<0) = m\delta_{m+n,0},\quad \sum_{m > 0} \frac 
{\pi(-m)\pi(m)}{m} = 1 - |k\>\<k|
\end{equation}
\vspace{-0.2in}
\begin{equation}
\label{eq3.4b}
{\tilde \pi}(m>0){\tilde \pi}(n<0) = m\delta_{m+n,0},\quad \sum_{m>0} 
\frac {{\tilde
\pi}(-m){\tilde \pi}(m)}{m} = 1 - |k\>\<k|.
\end{equation}
Together, Eqs. (3.3) and (3.4) form an infinite-dimensional version of a so-called
{\it symmetric Bose/Cuntz algebra} [4]. See Refs. [4-6] and the Appendix for further
applications of the tilde operators.

As for the fermions above, I will confine our discussion  here primarily to the
subalgebra (3.4a) of  the untilde operators, which is in fact an  infinite-dimensional
Cuntz algebra:
\setcounter{subsection}{5}
\setcounter{equation}{0}
\begin{equation}
\label{eq3.5a}
a(m) \equiv \frac {\pi(m)}{\sqrt{m}},\quad a^{\dag}(m) = \frac {\pi(-m)}{m},\quad m > 0
\end{equation}
\vspace{-0.2in}
\begin{equation}
\label{eq3.5b}
a(m)a^{\dag}(n) = \delta_{m,n},\quad \sum_{m > 0} a^{\dag}(m)a(m) = 1 - |k\>\<k|
\end{equation}
\vspace{-0.1in}
\begin{equation}
\label{eq3.5c}
a(m) |k\> = \<k| a^{\dag}(m) = 0
\end{equation}
\vspace{-0.1in}
\begin{equation}
\label{eq3.5d}
|w,k\> = a_w{}^{\dag}|k\>,\quad \<w',k'|w,k\> = \delta_{w',w}\delta_{k',k}.
\end{equation}
The reduced bosonic words $w$ are formed as in (2.7) with 
$b,b{}^{\dag} \rightarrow  a,a{}^{\dag}$.
I emphasize that, up to a relabeling, this Cuntz algebra is that same as we found
above for the reduced fermions -- so that all reduced states, fermionic or bosonic,
satisfy the same classical or Boltzmann statistics at large $N$.

Following the logic of Section 2, let us move quickly in this case to find the
explicit form of the  reduced ``Hamiltonian" $L_B(0)$ of the large $N$  system. This
trace class operator must satisfy the reduced commutation relation
\renewcommand{\theequation}{\thesection.\arabic{equation}}
\setcounter{equation}{5}
\begin{equation}
\label{eq3.6}
[L_B(0),\pi(m)] = -m\pi(m),\quad L_B{}^{\dag}(0) = L_B(0)
\end{equation}
which is the reduced image of the familiar commutator in the unreduced theory.
The solution of (3.6) is 
\renewcommand{\theequation}{\thesection.\arabic{subsection}\alph{equation}}
\setcounter{subsection}{7}
\setcounter{equation}{0}
\begin{equation}
\label{eq3.7a}
L_B(0) - \frac {\pi^2(0)}{2} = \sum_w a_w{}^{\dag} \Big{(}\sum_{m > 0} 
\pi(-m)\pi(m)\Big{)}a_w
\end{equation}
\vspace{-0.2in}
\begin{equation}
\label{eq3.7b}
(L_B(0) - \Delta(w,k))|w,k\> = \<w,k| (L_B(0)-\Delta(w,k)) = 0
\end{equation}
\vspace{-0.2in}
\begin{equation}
\label{eq3.7c}
\Delta(w,k) = \frac {k^2}{2} + \{w\}_B,\quad \{w\}_B \equiv \sum_{j=1}^n m_j
\end{equation}
where the weight $\{w\}_B$ is again the mode level of the basis state $|w,k\>$.  We will also
need the generalization of these results to the case of $D$ chiral adjoint bosons
\setcounter{subsection}{8}
\setcounter{equation}{0}
\begin{equation}
\label{eq3.8a}
\pi_i(m>0)\pi_j(n<0) = \delta_{ij}m\delta_{m+n,0},\quad i,j = 1\dots D
\end{equation}
\vspace{-0.2in}
\begin{equation}
\label{eq3.8b}
\sum_{i=1}^D \sum_{m>0} \frac {\pi_i(-m)\pi_i(m)}{m} = 1-|k\>\<k|,\quad 
\pi_i{}^{\dag}(m) =
\pi_i(-m)
\end{equation}
\vspace{-0.2in}
\begin{equation}
\label{eq3.8c}
(\pi_i(m\ge 0) - \delta_{m,0}k_i)|k\> = \<k|(\pi_i(m\le 0) - 
\delta_{m,0}k_i) = 0
\end{equation}
\vspace{-0.2in}
\begin{equation}
\label{eq3.8d}
L_B(0) - \frac {{\overrightarrow{\pi}}^2(0)}{2} = \sum_w a_w{}^{\dag} 
\sum_{i=1}^D
\sum_{m>0} \pi_i(-m)\pi_i(m)a_w
\end{equation}
\vspace{-0.2in}
\begin{equation}
\label{eq3.8e}
L_B{}^{\dag}(0) = L_B(0),\quad [L_B(0),\pi_i(m)] = -m\pi_i(m)
\end{equation}
where the words of the dressing in (3.8d) are now $w = (m_1,i_1...m_n,i_n)$.
With the substitution $k^2 \rightarrow \vec{k}^2$, the action of $L_B(0)$ on the basis states
is the same as that given in (3.7c).

We are now prepared to define and evaluate the Cuntz-algebraic
{\it reduced bosonic partition functions}
\setcounter{subsection}{9}
\setcounter{equation}{0}
\begin{eqnarray}
\label{eq3.9a}
Z_B(z) &\equiv & \mbox{Tr}_k(z^{L_B(0)}) = z^{\frac 
{{\overrightarrow{k}}^2}{2}} \sum_w
z^{\{w\}_B}\\
& =& z^{\frac {{\overrightarrow{k}}^2}{2}} \frac {1}{1-D \sum_{m > 0} 
z^m} = z^{\frac
{{\overrightarrow{k}}^2}{2}} \frac {1-z}{1-(D+1)z}\\
& =& z^{\frac {{\overrightarrow{k}}^2}{2}} \left( 1 + 
\sum_{n=1}^{\infty} d_n(D)z^n\right)
\end{eqnarray}
\vspace{-0.2in}
\begin{equation}
\label{eq3.9d}
d_n(D) = D[(D+1)^{n-1}],\quad n \ge 1
\end{equation}
which provide simpler examples of the phenomena discussed above for the
reduced fermions. In particular, one finds the radius of convergence $z_0(D)$,
the 2-d limiting temperature $\beta_0(D)^{-1}$ and the apparently universal
$O(\exp(an))$ growth in the number of states
\renewcommand{\theequation}{\thesection.\arabic{equation}}
\setcounter{equation}{9}
\begin{equation}
\label{eq3.10}
z_0(D) = \frac {1}{D+1} < 1,\quad \beta_0(D) = \ln(D+1),\quad d_n(D) \ {}_{\stackrel{\mbox{\Large $\sim$}}{n \gg 1}}\
e^{n\beta_0(D)}
\end{equation}
each of which is a consequence of the Boltzmann statistics associated to
the Cuntz algebra. We remind the reader that the conventional fermionic and
bosonic string partition functions show an infinite 2-d limiting temperature
associated to growth of order $\exp(b\sqrt{n})$.

\setcounter{subsection}{1}
\subsection{Reduced bosonic $\mathfrak{sl}(2)$}
\label{sec3.2}

Following the discussion above for the fermions I turn next to complete the
reduced bosonic $\mathfrak{sl}(2)$ generators $L_B(m)$, starting from the reduced commutation
relations
\begin{equation}
\label{eq3.11}
[L_B(l),\pi(n)] = -n\pi(n+l).
\end{equation}
Using these relations and the Cuntz algebra (3.4a), we find the consistency
relations
\begin{equation}
\label{eq3.12}
m\pi(m+l)\pi(n) + n\pi(m)\pi(n+l) = 0,\quad m > 0,\ n < 0
\end{equation}
and, as argued above on general grounds,  these relations are easily seen
to be inconsistent beyond the $\mathfrak{sl}(2)$ subalgebra. For example, the inconsistency 
\begin{equation}
\label{eq3.13}
\pi(m)\pi(1) = 0,\quad m > 0
\end{equation}
is obtained after using the Cuntz algebra for the case $l=2,n=-1$.

But there is also something new here. The choice $l=1,n=-1$ gives in the same way
the further relations
\begin{equation}
\label{eq3.14}
\pi(m)\pi(0) = 0,\quad m > 0
\end{equation}
which can only be satisfied by setting the reduced momenta to zero
\begin{equation}
\label{eq3.15}
\pi(0) = k = 0
\end{equation}
and keeping only the Cuntz module $a_w{}^{\dag}|0\>$ based on the reduced zero momentum
ground state $|0\>$. I do not have a deeper understanding of this phenomenon,
although it is presumably traceable to the rescaling of the unreduced momenta
which was necessary to obtain finite reduced momenta in the first place.

With the proviso (3.15), it is not difficult to solve the commutation 
relations (3.11) to obtain the explicit form
of the reduced bosonic $\mathfrak{sl}(2)$ generators
\renewcommand{\theequation}{\thesection.\arabic{subsection}\alph{equation}}
\setcounter{subsection}{16}
\setcounter{equation}{0}
\begin{equation}
\label{eq3.16a}
L_B(|m|\le 1) = \sum_w a_w{}^{\dag} B_B(m)a_w
\end{equation}
\vspace{-0.2in}
\begin{equation}
\label{eq3.16b}
B_B(m) = \sum_{n > 0} \pi(-n)\pi(n+m)
\end{equation}
and this agrees with our previous result for $L_B(0)$ when $\pi(0)=0$.

Continuing to compute with the Cuntz algebra, one verifies the additional
properties of the kernels
\setcounter{subsection}{17}
\setcounter{equation}{0}
\begin{equation}
\label{eq3.17a}
B_B(m)^{\dag} = B_B(-m),\quad B_B(m)|0\> = \<0|B_B(m) = 0
\end{equation}
\begin{equation}
\label{eq3.17b}
\pi(m>0)B_B(n) = -B_B(n)\pi(m<0) = m\pi(m+n)
\end{equation}
\begin{equation}
\label{eq3.17c}
[B_B(m),B_B(n)] = (m-n)B_B(m+n)
\end{equation}
\begin{equation}
\label{eq3.17d}
[L_B(m),B_B(n)] = (m-n)B_B(m+n)
\end{equation}
and finally one verifies the reduced $\mathfrak{sl}(2)$ algebra
\setcounter{subsection}{18}
\setcounter{equation}{0}
\begin{equation}
\label{eq3.18a}
[L_B(m),L_B(n)] = \sum_w a_w{}^{\dag} [L_B(m),B_B(n)]a_w = (m-n)L_B(m+n)
\end{equation}
\vspace{-0.2in}
\begin{equation}
\label{eq3.18b}
L_B(m)^{\dag} = L_B(-m),\quad L_B(m)|0\> = \<0|L_B(m) = 0.
\end{equation}
Note that the reduced bosonic ground state $|0\>$ is  $\mathfrak{sl}(2)$ invariant, as it is
in the unreduced bosonic theory.  To generalize these results to the 
Cuntz algebra (3.8a,b) of $D$ reduced bosons, use the kernels
\renewcommand{\theequation}{\thesection.\arabic{equation}}
\setcounter{equation}{18}
\begin{equation}
\label{eq3.19}
B_B(|m| \le 1) = \sum_{i=1}^D\sum_{n>0}\pi_i(-n)\pi_i(n+m)
\end{equation}
and construct the dressing with the  full words $(m_1,i_1...m_n,i_n)$ as above.

This concludes my  discussion of reduced chiral fermions and bosons separately,
and I turn next to  some applications which, in a variety of ways,  combine
reduced chiral fermions and bosons in the same system.

\section{Extended Affine Free Algebra}
\label{sec4}

\subsection{Short and long currents}
\label{sec4.1}

Returning to the reduced chiral fermions of Sec. 2, I want to discuss here
an extended affine free algebra which contains  both the affine free algebra
and a surprising new copy of the Cuntz algebra among its free subalgebras.  In
turn, all these  algebras are subalgebras of the algebra of all normal-ordered
products of Cuntz operators (see  Eqs. (88-90) of Ref. [8] with $a\rightarrow b$),
but we shall not need the explicit form of this much larger algebra. Although it is
straightforward to extend this construction to any number of  reduced fermions,
I will confine the discussion for simplicity to the case of a single reduced
fermion, and hence reduced level $\hat{x}=1$ of the affine free algebra. 

To begin,  let us decompose the free currents (2.13) into the following 
{\it short
currents} $J_{++}, J_{--}$ and {\it long currents} $J_{-+}$:
\renewcommand{\theequation}{\thesection.\arabic{subsection}\alph{equation}}
\setcounter{subsection}{1}
\setcounter{equation}{0}
\begin{equation}
\label{eq4.1a}
J(m \ge 0) = J_{++}(m \ge 0) + J_{-+}(m\ge 0)
\end{equation}
\begin{equation}
\label{eq4.1b}
J_{++}(m \ge 0) \equiv -\sum_{0<p<m} H(m-p)H(p)
\end{equation}
\vspace{-0.1in}
\begin{equation}
\label{eq4.1c}
J_{-+}(m \ge 0) \equiv -\sum_{p>m} H(m-p)H(p)
\end{equation}
\vskip 0.1 in
\setcounter{subsection}{2}
\setcounter{equation}{0}
\begin{equation}
\label{eq4.2a}
J(m \le 0) = J_{--}(m \le 0) + J_{-+}(m \le 0)
\end{equation}
\begin{equation}
\label{eq4.2b}
J_{--}(m \le 0) \equiv -\sum_{0<p<-m} H(-p)H(m+p)
\end{equation}
\vspace{-0.1in}
\begin{equation}
\label{eq4.2c}
J_{-+}(m\le 0) \equiv -\sum_{p>-m} H(-p)H(m+p).
\end{equation}
The short currents are a sum of a finite number of terms, and the subscripts
are such that, for example, $J_{-+}$  has negative (positive) reduced fermionic modes
on the left (right).

Simple properties of the short and long currents include:
\setcounter{subsection}{3}
\setcounter{equation}{0}
\begin{equation}
\label{eq4.3a}
J_{++}(m\ge 0)^{\dag} = J_{--}(-m),\quad J_{-+}(m\ge 0)^{\dag} = J_{-+}(-m)
\end{equation}
\begin{equation}
\label{eq4.3b}
J_{++}(0) = J_{--}(0) = 0,\quad J_{-+}(0) = J(0) = |0\>\<0| - 1
\end{equation}
\begin{equation}
\label{eq4.3c}
J_{++}(m \ge 0)|0\> = J_{-+}(m \ge 0)|0\> = J_{-+}(m\le 0)|0\> = 0
\end{equation}
\begin{equation}
\label{eq4.3d}
\<0|J_{--}(m \le 0) = \<0|J_{-+}(m\ge 0) = \<0|J_{-+}(m \ge 0) = 0.
\end{equation}
We will also need the relations of these currents with the reduced fermions
\setcounter{subsection}{4}
\setcounter{equation}{0}
\begin{equation}
\label{eq4.4a}
H(p>0)J_{--}(m \le 0) = J_{-+}(m \ge 0)H(p<0) = -\theta(p+m<0)H(p+m)
\end{equation}
\begin{equation}
\label{eq4.4b}
H(p>0)J_{-+}(m \le 0) = J_{++}(m\ge 0)H(p<0) = -\theta(p+m>0)H(p+m)
\end{equation}
which follow from the fermion Cuntz algebra (2.5a).

Using (4.4), we then obtain the {\it extended affine free algebra}
\setcounter{subsection}{5}
\setcounter{equation}{0}
\begin{equation}
\label{eq4.5a}
J_{++}(m \ge 0)J_{--}(n \le 0) = m\delta_{m+n,0}
\end{equation}
\vspace{-0.1in}
\begin{equation}
\label{eq4.5b}
J_{++}(m \ge 0)J_{-+}(n \le 0) = -\theta(m + n \ge 0)J_{++}(m+n)
\end{equation}
\begin{equation}
\label{eq4.5c}
J_{-+}(m \ge 0)J_{--}(n\le 0) = -\theta(m+n\le 0)J_{--}(m+n)
\end{equation}
\begin{equation}
\label{eq4.5d}
J_{-+}(m\ge 0)J_{-+}(n \le 0) = -J_{-+}(m+n)
\end{equation}
\begin{equation}
\label{eq4.5e}
J_{-+}(m \ge 0)J_{-+}(n \ge 0) = J_{-+}(m\le 0)J_{-+}(n \le 0) = -J_{-+}(m+n)
\end{equation}
whose properties I will discuss below. There are also three further relations
\setcounter{subsection}{6}
\setcounter{equation}{0}
\begin{equation}
\label{eq4.6a}
J_{++}(m\ge 0)J_{-+}(n \ge 0) = \sum_{0<p<m} H(m-p)H(p+n)
\end{equation}
\vspace{-0.1in}
\begin{equation}
\label{eq4.6b}
J_{-+}(m \le 0)J_{--}(n\le 0) = \sum_{0<p<-n} H(m-p)H(p+n)
\end{equation}
\vspace{-0.1in}
\begin{eqnarray}
\label{eq4.6c}
& &J_{-+}(m \le 0)J_{-+}(n\ge 0) = \sum_{p>-m} H(-p)H(m+n+p) \nonumber \\
& &\hskip 1 in = \sum_{p>0}
H(m-p)H(p+n).
\end{eqnarray}
which open into new normal-ordered fermion bilinears. This exhausts the relations
among the short and long currents, which are all of the form $ A_{x+} B_{-y}= C_{xy}$.

Let us discuss the {\it free subalgebras} of the extended affine free algebra (4.5). In
the first place, we see that the long currents $J_{-+}$ form a closed free subalgebra
of the extended affine free algebra. Second, the first four relations in (4.5) and
the definitions (4.2a),(4.3a) can be used to verify that the affine free algebra (2.16)
is also a free subalgebra of the extended affine free algebra. 

Finally note that the short currents form a third {\it short free subalgebra} which
is recognized as another infinite-dimensional ``bosonic" Cuntz algebra
\setcounter{subsection}{7}
\setcounter{equation}{0}
\begin{eqnarray}
\label{eq4.7a}
a_S(m) &\equiv& \frac {J_{++}(m>0)}{\sqrt{m}} = -\frac {1}{\sqrt{m}} 
\sum_{0<p<m} b(m-p)b(p) \\
a_S{}^{\dag}(m) &=&  \frac {J_{--}(-m)}{\sqrt{m}} = -\frac 
{1}{\sqrt{m}} \sum_{0<p<m}
b^{\dag}(p)b^{\dag}(m-p)
\end{eqnarray}
\vspace{-0.1in}
\begin{equation}
\label{eq4.7c}
a_S(m)a_S{}^{\dag}(n) = \delta_{m,n},\quad a_S(m)|0\> = \<0|a_S{}^{\dag}(m) = 0
\end{equation}
\begin{equation}
\label{eq4.7d}
J(m>0) = \sqrt{m} a_S(m) + J_{-+}(m > 0)
\end{equation}
\begin{equation}
\label{eq4.7e}
J(-m) = \sqrt{m} a_S{}^{\dag}(m) + J_{-+}(-m)
\end{equation}
whose generators $a_S,a_S{}^{\dag}$ we have constructed as {\it bilinears} in the original
fermionic Cuntz operators $b,b^{\dag}$. Since the bosonic Cuntz algebra (4.7c) is
isomorphic to the fermionic Cuntz algebra (2.6b),  the relations in (4.7a,b)
define what can be called an automorphism of the Cuntz algebra.

It is also instructive to rewrite some of the relations above in the Cuntz notation:
\setcounter{subsection}{8}
\setcounter{equation}{0}
\begin{equation}
\label{eq4.8a}
b(p)a_S{}^{\dag}(m) = -\theta(m>p) \frac {1}{\sqrt{m}} b^{\dag}(m-p)
\end{equation}
\vspace{-0.1in}
\begin{equation}
\label{eq4.8b}
a_S(m)b^{\dag}(p) = -\theta(m>p) \frac {1}{\sqrt{m}} b(m-p)
\end{equation}
\vspace{-0.1in}
\begin{equation}
\label{eq4.8c}
a_S(m)J_{-+}(-n) = -\theta(m\ge n) \sqrt{\frac {m-n}{m}} a_S(m-n),\quad n \ge 0
\end{equation}
\vspace{-0.1in}
\begin{equation}
\label{eq4.8d}
J_{-+}(m)a_S{}^{\dag}(n) = -\theta(n \ge m) \sqrt{\frac {n-m}{n}} 
a_S{}^{\dag}(n-m),\quad m
\ge 0.
\end{equation}
In this form we see clearly that the ``bosonic" Cuntz generators (short currents)
are raising and lowering operators for the the fermionic Cuntz generators, and
moreover that the long currents are raising and lowering operators for the bosonic
Cuntz generators.

Applications of the extended affine free algebra are discussed in the following
two subsections.

\setcounter{subsection}{1}
\subsection{Free-current form of the fermionic $\mathfrak{sl}(2)$ kernels}
\label{sec4.2}

I have mentioned above that the Virasoro operators of one unreduced adjoint fermion
are equal to those of the affine-Sugawara construction [9,12,15-17] on the unreduced fermionic
currents (2.9a) at invariant level $N$ of affine $\mathfrak{su}(N)$. Since our extra
$\mathfrak{u}(1)$ fermion
is negligable at large $N$, this suggests that it may be possible to find  what could
be called a {\it free affine-Sugawara construction} -- in which the reduced fermionic
$\mathfrak{sl}(2)$ generators (2.32) are re-expressed entirely in terms of free currents.

Indeed, using the extended affine free algebra (4.5), I have been able to find an
equivalent {\it free-current form of the kernels} $B_F(m)$ of the reduced fermionic
$\mathfrak{sl}(2)$ generators
\setcounter{subsection}{9}
\setcounter{equation}{0}
\begin{equation}
\label{eq4.9a}
B_F(0) = \sum_{p>0} pH(-p)H(p) = \sum_{m \ge 1} J_{-+}(-m)J_{-+}(m) + 
\frac {1}{2}
J_{-+}(0)^2
\end{equation}
\vspace{-0.1in}
\begin{equation}
\label{eq4.9b}
B_F(1) = \sum_{p>0} (p + \frac{1}{2})H(-p)H(p+1) = \sum_{m \ge 0} J_{-+}(-m)J_{-+}(m+1)
\end{equation}
\vspace{-0.1in}
\begin{equation}
\label{eq4.9c}
B_F(-1) = \sum_{p>0} (p + \frac{1}{2})H(-p-1)H(p) = \sum_{m \ge 0} 
J_{-+}(-m-1)J_{-+}(m)
\end{equation}
but I have not yet found a free-current form of the operation $\sum_w b_w{}^{\dag}(...)b_w$
which is necessary to dress the kernels.

\setcounter{subsection}{2}
\subsection{A free-algebraic coset construction}
\label{sec4.3}

The first examples of coset constructions were given implicitly in Ref.[9] and
explicitly in Ref. [12], and the general coset construction was given later
in Ref. [18]. In this subsection I will present a 
{\it free-algebraic coset construction}
based on the ``bosonic" or short free subalgebra of the extended affine free algebra.

To begin this discussion, it is convenient to define a unified form of the short
currents:
\setcounter{subsection}{10}
\setcounter{equation}{0}
\begin{equation}
\label{eq4.10a}
J_S(m) \equiv \theta(m>0)J_{++}(m) + \theta(m<0)J_{--}(m),\quad \forall\ 
m \in {\mathbb Z}
\end{equation}
\begin{equation}
\label{eq4.10b}
J_S(0) = 0,\quad J_S{}^{\dag}(m) = J_S(-m)
\end{equation}
\begin{equation}
\label{eq4.10c}
J_S(m > 0)J_S(n<0) = m\delta_{m+n,0}.
\end{equation}
Then we find with (2.27) that the long and short currents  are  both $(1,0)$
operators under the reduced fermionic $sl(2)$
\setcounter{subsection}{11}
\setcounter{equation}{0}
\begin{equation}
\label{eq4.11a}
[L_F(m),J_S(n)] = -nJ_S(n+m),\quad \forall\ n \in {\mathbb Z}
\end{equation}
\begin{equation}
\label{eq4.11b}
[L_F(m),J_{-+}(n)] = -nJ_{-+}(n),\quad \forall\ n \in {\mathbb Z}
\end{equation}
where the explicit form of $\{L_F(m)\}$ is given in Eq. (2.32). This is of course
consistent with  our earlier observation in (2.35) that the full free current
$J = J_{S}+J_{-+}$  is a $(1,0)$ operator.

Next, I can construct  a new {\it short reduced} $\mathfrak{sl(2)}$ using only the
short currents
\setcounter{subsection}{12}
\setcounter{equation}{0}
\begin{equation}
\label{eq4.12a}
L_S(m) \equiv \sum_w (a_S)_w{}^{\dag}\left( \sum_{n>0} 
J_S(-n)J_S(n+m)\right)(a_S)_w
\end{equation}
\vspace{-0.1in}
\begin{equation}
\label{eq4.12b}
[L_S(m),L_S(n)] = (m-n)L_S(m+n)
\end{equation}
\begin{equation}
\label{eq4.12c}
[L_S(m),J_S(n)] = -nJ_S(n+m)
\end{equation}
in complete analogy with the reduced bosonic $\mathfrak{sl}(2)$ in Eq. (3.16).  We do not
need to check these relations explicitly because the Cuntz algebra (4.10c) of
the short currents is isomorphic to the bosonic Cuntz algebra of the reduced
bosonic modes$\{\pi(m)\}$. Note however that the kernels of the short reduced
$\mathfrak{sl}(2)$
are {\it quartic} in the reduced fermion modes.

We are now ready for the free-algebraic coset construction-- in which we mod out
the reduced fermion theory by the short or bosonic free subalgebra (4.7) of the extended
affine free algebra (4.5). The key of course is that the short currents are $(1,0)$
operators under both sets $\{L_F(m)\}$ and $\{L_S(m)\}$ of reduced $\mathfrak{sl}(2)$
generators. Then the short currents commute with the  differences $\{L_{F/S}(m)\}$
of the generators
\setcounter{subsection}{13}
\setcounter{equation}{0}
\begin{eqnarray}
\label{eq4.13a}
L_{F/S}(m) &\equiv & L_F(m) - L_S(m),\quad |m| \le 1 \\
&=& \sum_w \left\{ b_w{}^{\dag} \sum_{p>0} (\frac{m}{2} + p)H(-p)H(p+m)b_w 
\right. \nonumber \\
& &\hskip 0.6 in \left. -
(a_S)_w{}^{\dag}
\sum_{n>0} J_S(-n)J_S(n+m)(a_S)_w\right\}
\end{eqnarray}
\vspace{-0.1in}
\begin{equation}
\label{eq4.13c}
[L_{F/S}(m),J_S(n)] = 0
\end{equation}
and we may  follow the familiar steps
\setcounter{subsection}{14}
\setcounter{equation}{0}
\begin{equation}
\label{eq4.14a}
[L_{F/S}(m),L_S(n)] = 0
\end{equation}
\begin{equation}
\label{eq4.14b}
[L_{F/S}(m),L_{F/S}(n)] = (m-n)L_{F/S}(m+n)
\end{equation}
to show that the operators $\{L_{F/S}(m)\}$ generate a new {\it reduced coset} $\mathfrak{sl}(2)$. So far as
I can tell, this free-algebraic coset construction has no analogue at finite values of $N$.

\section{Free-Algebraic Construction of $\mathfrak{osp}(1|2)$}
\label{sec5}

Historically, the next step was taken by Neveu and Schwarz [19], who noticed that
the idea of Ramond's superconformal symmetry [11] could be combined  with
half-integral moded world sheet fermions of the  Bardakci- Halpern type to find
the so-called NS superconformal construction. In this section I will consider the
large N limit of the simplest matrix generalization of the NS construction, namely
that composed of  one chiral boson-fermion pair in the adjoint. This construction
uses our previous knowledge of reduced chiral fermions and reduced chiral bosons,
as well as the technology developed for the construction of various reduced
supersymmetries in Ref. [4].

In this case, we know from the discussion above that only the reduced $\mathfrak{osp}(1|2)$
subalgebra of the superconformal algebra
\setcounter{subsection}{1}
\setcounter{equation}{0}
\begin{equation}
\label{eq5.1a}
\{G(r),r = \pm 1/2;\,\,\, L(m),m=0,\pm 1\}
\end{equation}
\begin{equation}
\label{eq5.1b}
[G(r),\pi(m)] = -mH(r+m),\quad [G(r),H(p)]_+ = \pi(r+p)
\end{equation}
\begin{equation}
\label{eq5.1c}
[G(r),G(s)]_+ = 2L(r+s),\quad [L(m),L(n)] = (m-n)L(m+n)
\end{equation}
\begin{equation}
\label{eq5.1d}
[L(m),\pi(n)] = -n\pi(n+m),\quad [L(m),H(p)] = -(\frac{m}{2} + p)H(p+m)
\end{equation}
\begin{equation}
\label{eq5.1e}
[L(m),G(r)] = (\frac{m}{2} - r)G(r+m)
\end{equation}
will be well defined in the large $N$ limit. The reduced  fermion modes $\{H(p)\}$ and
reduced boson modes $\{\pi(m)\}$ with $\pi(0)= 0$ satisfy: 
\setcounter{subsection}{2}
\setcounter{equation}{0}
\begin{equation}
\label{eq5.2a}
\pi(m>0)\pi(n<0) = m\delta_{m+n,0},\quad H(p>0)H(q<0) = \delta_{p+q,0}
\end{equation}
\begin{equation}
\label{eq5.2b}
\pi(m>0)H(p<0) = H(p>0)\pi(m<0) = 0
\end{equation}
\vspace{-0.1in}
\begin{equation}
\label{eq5.2c}
\sum_{m>0} \frac {1}{m} \pi(-m)\pi(m) + \sum_{p>0} H(-p)H(p) = 1 - |0\>\<0|
\end{equation}
\vspace{-0.1in}
\begin{equation}
\label{eq5.2d}
\pi(m>0)|0\> = H(p>0)|0\> = \<0|\pi(m<0) = \<0|H(p<0) = 0.
\end{equation}
Derivation of such systems from the unreduced theory is discussed in 
Ref.[4]. As
an  important intermediate step one also obtains a tilde copy of each of the
relations in (5.2) plus the following relations between the tilde and untilde
operators
\setcounter{subsection}{3}
\setcounter{equation}{0}
\begin{equation}
\label{eq5.3a}
[\pi(m),{\tilde \pi}(n)] = \delta_{m+n,0} |0\>\<0|
\end{equation}
\begin{equation}
\label{eq5.3b}
[H(p),{\tilde H}(q)]_+ = \delta_{p+q,0} |0\>\<0|
\end{equation}
\begin{equation}
\label{eq5.3c}
[\pi(m),{\tilde H}(p)] = [{\tilde \pi}(m),H(p)] = 0
\end{equation}
and all these relations together form an infinite-dimensional version of a so-called
{\it symmetric Bose/Fermi/Cuntz algebra} [4]. Again, the tilde operators will not be needed
in the discussion below.

Note that the relations (5.2) can be written as an infinite-dimensional {\it Cuntz
superalgebra} [4]
\setcounter{subsection}{4}
\setcounter{equation}{0}
\begin{equation}
\label{eq5.4a}
S(m) \equiv \bmatrix a(m) \\ b(m-\frac{1}{2}) \endbmatrix,\quad S^{\dag}(m) =
[a^{\dag}(m),b^{\dag}(m-\frac{1}{2})],\quad m = 1,2,\dots
\end{equation}
\begin{equation}
\label{eq5.4b}
S_i(m)S_j{}^{\dag}(n) = \delta_{ij}\delta_{m,n},\quad i,j = 1,2
\end{equation}
\begin{equation}
\label{eq5.4c}
\sum_m S^{\dag}(m)S(m) = 1 - |0\>\<0|,\quad S(m)|0\> = \<0|S^{\dag}(m) = 0
\end{equation}
in which the reduced fermions and reduced bosons are on equal footing.  This
gives us the useful {\it superword} notation
\setcounter{subsection}{5}
\setcounter{equation}{0}
\begin{equation}
\label{eq5.5a}
S_w = S_{i_1}(m_1) \dots S_{i_n}(m_n),\quad w = (i_1m_1,\dots,i_nm_n)
\end{equation}
\begin{equation}
\label{eq5.5b}
|w\> = S_w{}^{\dag}|0\> = S_{i_n}{}^{\dag}(m_n)\dots S_{i_1}{}^{\dag}(m_1)|0\>
\end{equation}
for describing the general basis state in the combined system.

The total reduced $\mathfrak{sl}(2)$ operators of the system are obtained as above by solving
the simultaneous reduced algebraic relations  in (5.1d). Following the discussion
of Ref. [4], one finds that the  kernels of these operators are additive, but that
the dressing of the  kernels  must be a {\it superdressing} 
\setcounter{subsection}{6}
\setcounter{equation}{0}
\begin{equation}
\label{eq5.6a}
L(m) = \sum_w S_w{}^{\dag}B(m)S_w,\quad |m| \le 1
\end{equation}
\begin{equation}
\label{eq5.6b}
B(m) = B_F(m) + B_B(m)
\end{equation}
\begin{equation}
\label{eq5.6c}
L(m)|0\> =\<0|L(m) = 0
\end{equation}
constructed with the generators $S,S^{\dag}$ of the Cuntz superalgebra (5.4). The  fermionic and
bosonic kernels $B_F$ and $B_B$ are given respectively in Eqs. (2.32b) and  (3.16b).
It is not difficult to check as above that these operators generate the reduced
$\mathfrak{sl}(2)$ in
(5.1c), and moreover that the Fermi and Bose terms of $L(m)$
\renewcommand{\theequation}{\thesection.\arabic{equation}}
\setcounter{equation}{6}
\begin{equation}
\label{eq5.7}
\sum_w S_w{}^{\dag} B_F(m)S_w,\quad \sum_w S_w{}^{\dag}B_B(m)S_w
\end{equation}
commute to generate a reduced $\mathfrak{sl}(2)\oplus\mathfrak{sl}(2)$.

The strategy to find the realization of the {\it reduced supercharges}  $\{G(r)\}$ is
to solve the simultaneous algebraic relations  in (5.1b), following Ref. [4].
The answer is
\renewcommand{\theequation}{\thesection.\arabic{subsection}\alph{equation}}
\setcounter{subsection}{8}
\setcounter{equation}{0}
\begin{equation}
\label{eq5.8a}
G(r) = \sum_w (S\tau_3)_w{}^{\dag} K(r)S_w = \sum_w S_w{}^{\dag}K(r)(\tau_3S)_w
\end{equation}
\vspace{-0.1in}
\begin{equation}
\label{eq5.8b}
K(r) = \sum_{m>0} \{H(r-m)\pi(m) + \pi(-m)H(m+r)\}
\end{equation}
\vspace{-0.1in}
\begin{equation}
\label{eq5.8c}
K(r)^{\dag} = K(-r),\quad G(r)^{\dag} = G(-r)
\end{equation}
\begin{equation}
\label{eq5.8d}
G(r)|0\> = \<0|G(r) = 0
\end{equation}
where $\tau_3$ is the third Pauli matrix, which operates in the $2\times2$ space of the
Cuntz superalgebra. More explicitly, this factor means that there is an extra
minus sign associated with each reduced fermion pair in the superdressing
\renewcommand{\theequation}{\thesection.\arabic{equation}}
\setcounter{equation}{8}
\begin{equation}
\label{eq5.9}
L(m) = B(m) + \sum_{n>0} a^{\dag}(n)B(m)a(n) - \sum_{p>0} 
b^{\dag}(p)B(m)b(p) + \dots .
\end{equation}
To check the anticommutation relations among the reduced supercharges, we need
the identities
\begin{equation}
\label{eq5.10}
[G(r),K(s)]_+ = 2B(r+s),\quad [L(m),K(r)] = (\frac{m}{2} - r)K(r+m)
\end{equation}
which follow straightforwardly from (5.1b,d). Then one finds for example that
\begin{equation}
\label{eq5.11}
[G(r),G(s)]_+ = \sum_w S_w{}^{\dag}[G(r),K(s)]_+ S_w = 2L(r+s)
\end{equation}
and similarly for the commutator in (5.1e).

\section{Exotic Constructions at Large $N$}
\label{sec6}

Using the vertex operator [20] of the bosonic string, the next developments in
conformal field theory were the vertex operator construction of fermions [21-23]
and the vertex operator construction of  level one of  $\mathfrak{su}(N)$ [21-23], followed
some years later by the vertex operator construction of level one of simply laced $g$
[24]. Qualitatively, we do not expect to find  such constructions directly in the
large $N$ bosonic systems because chiral  adjoint matter lives at unreduced affine
levels which are multiples of $N$. Quantitatively, it appears at first that we are
denied any analogue of the vertex operators by the fact that the Cuntz operators
do not satisfy commutation relations. 

But  as we shall see below, one can use the bosonic Cuntz algebra to construct a
new set of  creation and annihilation operators whose algebra, although not canonical,
allows  us to find new {\it free-algebraic vertex operators} and constructions.

\subsection{Dressed Cuntz operators}
\label{sec6.1}

We return in this subsection to the infinite-dimensional bosonic Cuntz algebra
\renewcommand{\theequation}{\thesection.\arabic{subsection}\alph{equation}}
\setcounter{subsection}{1}
\setcounter{equation}{0}
\begin{equation}
\label{eq6.1a}
a(m)a^{\dag}(n) = \delta_{m,n},\quad \sum_{m>0} a^{\dag}(m)a(m) = 1 - |0\>\<0|
\end{equation}
\vspace{-0.1in}
\begin{equation}
\label{eq6.1b}
a(m)|0\> = \<0| a^{\dag}(m) = 0,\quad m = 1,2,\dots
\end{equation}
which governs (see Sec.3) the large $N$ limit of the conformal field theory of
one chiral adjoint boson. The results of this subsection hold as well for either a) any
reduced bosonic sector k (with $|0\>\rightarrow|k\>$) or b) any finite
dimensional Cuntz algebra.

In terms of these operators, let us define the following
new {\it dressed Cuntz operators}:
\setcounter{subsection}{2}
\setcounter{equation}{0}
\begin{equation}
\label{eq6.2a}
A(m) \equiv \sum_w a_w{}^{\dag} a(m)a_w,\quad A^{\dag}(m) = \sum_w a_w{}^{\dag}
a^{\dag}(m)a_w
\end{equation}
\vspace{-0.1in}
\begin{equation}
\label{eq6.2b}
A(m)|0\> = \<0|A^{\dag}(m) = 0.
\end{equation}
The reader will recall that I have used this  Cuntz dressing $\sum_w a_w{}^{\dag}(...)a_w$
a number of times above in the construction of reduced trace class operators 
(see also Refs.[4,8]), but this is the first time it has been used to dress reduced
densities such as the Cuntz operators themselves. It will be helpful to have  a few
examples of the action of the dressed operators on the ground state
\setcounter{subsection}{3}
\setcounter{equation}{0}
\begin{equation}
\label{eq6.3a}
A^{\dag}(m)|0\> = a^{\dag}(m)|0\>,\quad \<0|A(m) = \<0|a(m)
\end{equation}
\begin{equation}
\label{eq6.3b}
A^{\dag}(m)A^{\dag}(n)|0\> = A^{\dag}(m)a^{\dag}(n)|0\>
\end{equation}
\begin{equation}
\label{eq6.3c}
= (a^{\dag}(m)a^{\dag}(n) + a^{\dag}(n)a^{\dag}(m))|0\>
\end{equation}
which are easily checked from the definitions in (6.2).

Using the lemma (2.31) with $b\rightarrow a$, we find after some algebra that the dressed
Cuntz  operators satisfy the {\it commutation relations}:
\setcounter{subsection}{4}
\setcounter{equation}{0}
\begin{equation}
\label{eq6.4a}
[a(m),A^{\dag}(n)] = [A(m),a^{\dag}(n)] = \delta_{m,n}
\end{equation}
\begin{equation}
\label{eq6.4b}
[A(m),A(n)] = [A^{\dag}(m),A^{\dag}(n)] = 0.
\end{equation}
These commutation relations tell us that the dressed Cuntz operators are ``canonical"
to  the original Cuntz operators, and moreover that all states created by the dressed
creation operators
\renewcommand{\theequation}{\thesection.\arabic{equation}}
\setcounter{equation}{4}
\begin{equation}
\label{eq6.5}
|\mbox{Bose},W\> = A_W{}^{\dag}|0\>,\quad W = (m_1,\dots,m_n)
\end{equation}
are {\it Bose symmetric} in the letters of $W$, as illustrated in (6.3).

In spite of the simple commutation relations (6.4), the pairs  $(a, A^{\dag})$ and/or
$( A, a^{\dag})$ are by no means canonical in the ordinary sense. This is clear already
because $a$ and $a^{\dag}$ satisfy only the Cuntz algebra (6.1), and, moreover,  I find
after some algebra the following additional commutation relations:
\renewcommand{\theequation}{\thesection.\arabic{subsection}\alph{equation}}
\setcounter{subsection}{6}
\setcounter{equation}{0}
\begin{equation}
\label{eq6.6a}
[a(m),A(n)] = a(m)a(n),\quad [A^{\dag}(m),a^{\dag}(n)] = a^{\dag}(m)a^{\dag}(n)
\end{equation}
\begin{equation}
\label{eq6.6b}
[A(m),A^{\dag}(n)] = \sum_w a_w{}^{\dag}(\delta_{mn} + a^{\dag}(n)a(m))a_w.
\end{equation}
Verification of the last relation in (6.6) is somewhat involved, and the reader
may find helpful the following intermediate steps
\setcounter{subsection}{7}
\setcounter{equation}{0}
\begin{equation}
\label{eq6.7a}
A^{\dag}(m) = a^{\dag}(m) + \sum_{n>0} a^{\dag}(n)A^{\dag}(m)a(n)
\end{equation}
\vspace{-0.1in}
\begin{equation}
\label{eq6.7b}
[A(m),A^{\dag}(n)] = \delta_{m,n} + a^{\dag}(n)a(m) + \sum_{l>0}
a^{\dag}(l)[A(m),A^{\dag}(n)]a(l)
\end{equation}
\vspace{-0.1in}
\begin{equation}
\label{eq6.7c}
A = B + \sum_{l > 0} a^{\dag}(l)Aa(l) = \sum_w a_w{}^{\dag}Ba_w,\quad \forall\ B.
\end{equation}
Finally, I have checked all the Jacobi identities for the new algebra (6.4),(6.6).

\setcounter{subsection}{1}
\subsection{Quasi-fermionic excitations in the large $N$ Bose system}
\label{sec6.2}

In this subsection, I will use the new dressed bosonic Cuntz operators (6.2) to
construct a free-algebraic analogue of the vertex operator construction of one
complex world-sheet fermion.

To begin this discussion, I will  revert to our earlier form (see Subs. (3.1)) of
the reduced bosonic theory  in which we allowed the zero-mode $\pi(0)$ of the reduced
boson to be non-zero. Recall that in this case the reduced theory does not possess
a full set of reduced $\mathfrak{sl}(2)$ generators, but only the reduced ``Hamiltonian" $L_B(0)$
in (3.6). Then, following the conventional intuition,  we may consider the states
\setcounter{subsection}{8}
\setcounter{equation}{0}
\begin{equation}
\label{eq6.8a}
|BH\>_{\pm} = |k=\pm 1\>,\quad |CR\>_{\pm} = |k = \pm 1/2\>
\end{equation}
\begin{equation}
\label{eq6.8b}
L_B(0)|BH\>_{\pm} = \frac {1}{2} |BH\>_{\pm},\quad L_B(0)|CR\>_{\pm} = \frac {1}{8}
|CR\>_{\pm}
\end{equation}
as candidates for  Bardakci-Halpern (BH) fermion/antifermion states  and complex
Ramond (CR) ground states in the reduced bosonic theory.

To see if this interpretation makes sense, we need a local description of these
excitations. In particular, let us consider using the dressed creation operators
$\{A^{\dag}(m)\}$ and the Cuntz annihilation operators $\{a(m)\}$  to construct the following
{\it free-algebraic analogue} of the conventional bosonized form of a complex  chiral
world-sheet fermion:
\setcounter{subsection}{9}
\setcounter{equation}{0}
\begin{equation}
\label{eq6.9a}
{\bar \psi}(z) \equiv e^{iq} e^{\pi(0)\ln z} e^{\sum_{n>0} \frac
{A^{\dag}(n)}{\sqrt{n}}z^n} e^{-\sum_{n>0} \frac {a(n)}{\sqrt{n}}z^{-n}}
\end{equation}
\begin{equation}
\label{eq6.9b}
\psi(z) \equiv e^{-iq} e^{-\pi(0)\ln z} e^{-\sum_{n>0} \frac 
{A^{\dag}(n)}{\sqrt{n}}z^n}
e^{\sum_{n>0} \frac {a(n)}{\sqrt{n}} z^{-n}}
\end{equation}
\begin{equation}
\label{eq6.9c}
[a(m),A^{\dag}(n)] = \delta_{m,n},\quad [A^{\dag}(m),A^{\dag}(n)] = 0,\quad [q,\pi(0)] = i
\end{equation}
\begin{equation}
\label{eq6.9d}
a(m)|k\> = \<k|A^{\dag}(m) = 0,\quad |k\> = e^{ikq}|0\>.
\end{equation}
Here I have also introduced a canonical coordinate q which commutes with all
operators except the zero mode $\pi(0)$.  As a consequence, the operators $\bar{\psi},\psi$
and all the relations in (6.9c-d) look quite ordinary, except that there are no
relations among the Cuntz annihilation operators $\{a(m)\}$.

These free-algebraic vertex operators are constructed to have many of the desirable
properties of ordinary vertex operators. Note first that $\bar{\psi},\psi$ are interpolating
fields for the candidate BH states in (6.8)
\renewcommand{\theequation}{\thesection.\arabic{equation}}
\setcounter{equation}{9}
\begin{equation}
\label{eq6.10}
{\bar \psi}(0)|0\> = |k=1\>,\quad \psi(0)|0\> = |k=-1\>
\end{equation}
and  multiple applications of $\bar{\psi},\psi$ on the reduced ground state $|0\>$ generates an
integer-valued lattice of momenta and hence a ``BH sector" in which the new operators
are half-integral moded:
\renewcommand{\theequation}{\thesection.\arabic{subsection}\alph{equation}}
\setcounter{subsection}{11}
\setcounter{equation}{0}
\begin{equation}
\label{eq6.11a}
{\mathcal O}(z) = \sum_{p \in {\mathbb Z}+1/2} {\mathcal 
O}(p)z^{-p-1/2},\quad {\mathcal O} =
{\bar \psi},\psi
\end{equation}
\begin{equation}
\label{eq6.11b}
{\bar \psi}(-1/2)|0\> = |k=1\>,\ \psi(-1/2)|0\> = |k=-1\>.
\end{equation}
Similarly, $\bar{\psi},\psi$ generate a ``CR sector" when acting on the candidate CR 
ground states in (6.8) . Second, it is not difficult to verify the following operator
identities
\setcounter{subsection}{12}
\setcounter{equation}{0}
\begin{equation}
\label{eq6.12a}
[L_B(0),a(m)] = -ma(m),\quad [L_B(0),A^{\dag}(m)] = mA^{\dag}(m)
\end{equation}
\begin{equation}
\label{eq6.12b}
(z\partial_z + 1/2){\mathcal O}(z) = [L_B(0),{\mathcal O}(z)],\quad 
{\mathcal O} = {\bar
\psi},\psi
\end{equation}
which tell us that $\bar{\psi},\psi$ boost with "conformal weight" $1/2$, as expected
for chiral world-sheet fermions.

But we must also study the algebra of $\bar{\psi},\psi$. Using
the commutators in (6.9c),
one may compute in the usual way the exact operator products
\setcounter{subsection}{13}
\setcounter{equation}{0}
\begin{equation}
\label{eq6.13a}
{\bar \psi}(z)\psi(\omega) = \frac {1}{z-\omega} (z/\omega)^{\pi(0)} 
e^{\sum_{n>0} \frac
{A^{\dag}(n)}{\sqrt{n}} (z^n-\omega^n)} e^{-\sum_{n>0} \frac 
{a(n)}{\sqrt{n}} z^{-n}}
e^{\sum_{n>0} \frac {a(n)}{\sqrt{n}} \omega^{-n}}
\end{equation}
\vspace{-0.1in}
\begin{equation}
\label{eq6.13b}
\psi(\omega){\bar \psi}(z) = -\frac {1}{z-\omega} (z/\omega)^{\pi(0)} 
e^{\sum_{n>0} \frac
{A^{\dag}(n)}{\sqrt{n}} (z^n-\omega^n)} e^{\sum_{n>0} \frac 
{a(n)}{\sqrt{n}} \omega^{-n}}
e^{-\sum_{n>0} \frac {a(n)}{\sqrt{n}} z^{-n}}
\end{equation}
for $|z|>|\omega|$ and $|\omega|>|z|$ respectively. Comparing the right sides to the conventional result, the only
difference is that, because there are no relations among the Cuntz operators, we
cannot combine or commute the last two  `$a$ factors' in any simple way.   This
difference does not contribute, however, to the leading term of the operator product
expansions
\setcounter{subsection}{14}
\setcounter{equation}{0}
\begin{eqnarray}
\label{eq6.14a}
{\bar \psi}(z)\psi(\omega) &=& \frac {1}{z-\omega} + O(z-\omega)^0 \\
&\simeq & -\psi(\omega){\bar \psi}(z)
\end{eqnarray}
\vspace{-0.1in}
\begin{equation}
\label{eq6.14c}
\psi(z)\psi(\omega) = {\bar \psi}(z){\bar \psi}(\omega) = O(z-\omega)^0
\end{equation}
where $\simeq$ means analytic continuation. These familiar relations tempt us to conclude
that our new operators $\bar{\psi},\psi$ are ordinary complex chiral fermion fields.

But such a conclusion would be premature (and incorrect). The reason is that the right
sides of  the exact relations (6.13a,b) are {\it not the same} -- even by analytic
continuation -- which blocks the familiar contour trick  needed to obtain the mode
algebra from this system!

The operator products (6.13) can however be written in the following more useful form
\setcounter{subsection}{15}
\setcounter{equation}{0}
\begin{eqnarray}
\label{eq6.15a}
{\bar \psi}(z)\psi(\omega) &=& \frac {1}{z-\omega} (z/\omega)^{\pi(0)} 
e^{\sum_{n>0} \frac
{A^{\dag}(n)}{\sqrt{n}} (z^n-\omega^n)} \Omega^{-1}(z)\Omega(\omega)\\
&\simeq & -\psi(\omega){\bar \psi}(z)R(z,\omega)
\end{eqnarray}
\vspace{-0.1in}
\begin{equation}
\label{eq6.15c}
\Omega(z) \equiv e^{\sum_{n>0} \frac {a(n)}{\sqrt{n}} z^{-n}},\quad
 R(z,\omega) \equiv \Omega(z)\Omega^{-1}(\omega)\Omega^{-1}(z)\Omega(\omega)
\end{equation}
where $R(z,w)$ is the {\it commutant} of the operator $\Omega$. In this form, we can use for
example the BH mode expansions (6.11) and the usual contour trick to obtain the
commutant-dependent mode relations
\setcounter{subsection}{16}
\setcounter{equation}{0}
\begin{equation}
\label{eq6.16a}
{\bar \psi}(p)\psi(q) + \sum_{m,n=0}^{\infty} \psi(q-n){\bar 
\psi}(p-m)R_{mn}(a) =
\delta_{q+p,0},\quad p,q \in {\mathbb Z} + 1/2
\end{equation}
\vspace{-0.2in}
\begin{equation}
\label{eq6.16b}
R(z,\omega) = \sum_{m,n=0}^{\infty} R_{mn}(a)z^{-m}\omega^{-n}
\end{equation}
in the BH sector, and similarly for the CR sector with integer $p$ and $q$. Using
heat kernel methods, it is straightforward to work out the explicit form of the
modes $R_{mn}(a)$ of the commutant, beginning with 
\renewcommand{\theequation}{\thesection.\arabic{equation}}
\setcounter{equation}{16}
\begin{equation}
\label{eq6.17}
R_{00}(a) = 1,\quad R_{01}(a) = R_{10}(a) = 0,\dots
\end{equation}
but  I will confine myself here to some remarks about the full commutant. 

Using  Eqs. (6.9c) and (6.9d), one finds after some algebra that the commutant 
satisfies the simple relations
\renewcommand{\theequation}{\thesection.\arabic{subsection}\alph{equation}}
\setcounter{subsection}{18}
\setcounter{equation}{0}
\begin{equation}
\label{eq6.18a}
[A^{\dag}(m),R(z,\omega)] = 0,\quad R(z,\omega)|k\> = |k\>
\end{equation}
\begin{equation}
\label{eq6.18b}
R(z,\omega)A_W{}^{\dag}|k\> = A_W{}^{\dag}|k\>
\end{equation}
and so it follows that $\bar{\psi},\psi$ behave as ordinary chiral complex fermions
when acting on the Bose symmetric states  $A_W{}^{\dag}|k\>$:
\setcounter{subsection}{19}
\setcounter{equation}{0}
\begin{equation}
\label{eq6.19a}
([{\bar \psi}(p),\psi(q)]_+ - \delta_{p+q,0})A_W{}^{\dag}|k\> = 0
\end{equation}
\begin{equation}
\label{eq6.19b}
[\psi(p),\psi(q)]_+ A_W{}^{\dag}|k\>
= [{\bar \psi}(p),{\bar \psi}(q)]_+A_W{}^{\dag}|k\> = 0
\end{equation}
\begin{equation}
\label{eq6.19c}
p,q \in \begin{cases}
{\mathbb Z} + 1/2 &\mbox{for $BH$} \\
{\mathbb Z} &\mbox{for $CR$}
\end{cases} ,\quad k \in \begin{cases}
{\mathbb Z} &\mbox{for $BH$} \\
{\mathbb Z} + 1/2 &\mbox{for $CR$}
\end{cases} .
\end{equation}
On the other hand, we cannot expect to find this simplification in general because
the Bose symmetric states are not generic in the Cuntz Hilbert space.

Recall that the set$\{ A_W{}^{\dag}|k>\}$ includes the low-lying states
\renewcommand{\theequation}{\thesection.\arabic{equation}}
\setcounter{equation}{19}
\begin{equation}
\label{eq6.20}
\{|k\>,\,\,a^{\dag}(m)|k\>,\,\,[a^{\dag}(m),a^{\dag}(n)]_+|k\>,\dots\}
\end{equation}
so that the lowest states with non-trivial commutant $R(z,w)$ are the antisymmetric
states $[a^{\dag}(m),a^{\dag}(n)]|k\>$.  As an example, I have worked out the  explicit form
taken by the mode relations (6.16a) 
\renewcommand{\theequation}{\thesection.\arabic{subsection}\alph{equation}}
\setcounter{subsection}{21}
\setcounter{equation}{0}
\begin{equation}
\label{eq6.21a}
\{R(z,\omega) - 1\}[a^{\dag}(m),a^{\dag}(n)]|k\> = \frac {2}{\sqrt{mn}}
(z^{-m}\omega^{-n}-\omega^{-m}z^{-n})
\end{equation}
\begin{eqnarray}
\label{eq6.21b}
& &\{[{\bar \psi}(p),\psi(q)]_+ - 
\delta_{p+q,0}\}[a^{\dag}(m),a^{\dag}(n)]|k\> \nonumber
\\
& &\hskip 0.25 in =
\frac {2}{\sqrt{mn}} \{\psi(q-m){\bar \psi}(p-n)-\psi(q-n){\bar 
\psi}(p-m)\}|k\>
\end{eqnarray}
when acting on these states.

In this subsection I have begun the discussion of an unexpected new excitation
or soliton in the large $N$ limit of the conformal field theory of a single
chiral adjoint boson. One might call the new excitation a complex chiral
{\it quasi-fermion} because it acts like a complex chiral fermion in a certain sector
of the Cuntz Hilbert space, and we have found both a half-integral moded BH sector
and an integral-moded  CR sector for the chiral quasi-fermion. Finally, it is not
difficult to find the {\it free-algebraic intertwiners} or spin fields
\setcounter{subsection}{22}
\setcounter{equation}{0}
\begin{equation}
\label{eq6.22a}
\sigma_{\pm}(z) = e^{\pm iq/2} e^{\pm \frac{1}{2} \pi(0)\ln z} e^{\pm \frac 
{1}{2} \sum_{n>0} \frac
{A^{\dag}(n)}{\sqrt{n}} z^n} e^{\mp \frac{1}{2} \sum_{n>0} \frac 
{a(n)}{\sqrt{n}} z^{-n}}
\end{equation}
\begin{equation}
\label{eq6.22b}
\sigma_{\pm}(0)|0\> = |k=\pm 1/2\>
\end{equation}
which connect the BH and the CR sectors of the chiral quasi-fermion.

\setcounter{subsection}{2}
\subsection{Free-algebraic vertex operators}
\label{sec6.3}

The free-algebraic vertex operators in Eqs.(6.9),(6.22) are easily generalized
as follows. 

We know that the large $N$ limit  of $D$ chiral adjoint bosons is described by a set
of $D$ reduced chiral boson modes $\{\pi_i(m)\}$ which satisfy the bosonic Cuntz algebra
(3.8a). Following Subsec.(6.1), we may then  construct $D$ pairs  of dressed creation
and annihilation operators
\setcounter{subsection}{23}
\setcounter{equation}{0}
\begin{equation}
\label{eq6.23a}
A_i(m) = \sum_\omega a_\omega{}^{\dag} a_i(m)a_\omega,\quad 
A_i{}^{\dag}(m) = \sum_\omega a_\omega{}^{\dag}
a_i{}^{\dag}(m)a_\omega
\end{equation}
\vspace{-0.2in}
\begin{equation}
\label{eq6.23b}
i = 1\dots D,\ m \in {\mathbb Z}
\end{equation}
\begin{equation}
\label{eq6.23c}
[A_i(m),a_j{}^{\dag}(n)] = [a_i(m),A_j{}^{\dag}(n)] = \delta_{ij}\delta_{m,n}
\end{equation}
\begin{equation}
\label{eq6.23d}
[A_i(m),A_j(n)] = [A_i{}^{\dag}(m),A_j{}^{\dag}(n)] = 0
\end{equation}
\begin{equation}
\label{eq6.23e}
[a_i(m),A_j(n)] = a_i(m)a_j(n),\quad [A_i{}^{\dag}(m),a_j{}^{\dag}(n)] =
a_i{}^{\dag}(m)a_j{}^{\dag}(n)
\end{equation}
\begin{eqnarray}
\label{eq6.23f}
[A_i(m),A_j{}^{\dag}(n)] &=& \sum_\omega 
a_\omega{}^{\dag}(\delta_{ij}\delta_{mn} +
a_j{}^{\dag}(n)a_i(m))a_\omega \\
&=& \sum_\omega a_\omega{}^{\dag} [a_i(m), a_j{}^{\dag}(n)]_{+} a_\omega 
\end{eqnarray}
\vspace{-0.2in}
\begin{equation}
\label{eq6.23h}
A_i(m)|k\> = \<k|A_i{}^{\dag}(m) = 0
\end{equation}
where the dressing includes the appropriate sums over $i$. The states constructed
from the dressed creation operators
\renewcommand{\theequation}{\thesection.\arabic{equation}}
\setcounter{equation}{23}
\begin{equation}
\label{eq6.24}
A_W{}^{\dag}|k\> = A_{i_n}{}^{\dag}(m_n)\dots A_{i_1}{}^{\dag}(m_1)|k\>,\quad W =
\{i_1m_1,\dots,i_nm_n\}
\end{equation}
are Bose symmetric, as above, in the letters of $W$. 

Then we may consider the more general {\it free-algebraic vertex operator}
\renewcommand{\theequation}{\thesection.\arabic{subsection}\alph{equation}}
\setcounter{subsection}{25}
\setcounter{equation}{0}
\begin{equation}
\label{eq6.25a}
U(\alpha,z) \equiv e^{i\alpha\cdot q} z^{\alpha\cdot\pi(0)} 
e^{\alpha\cdot\sum_{n>0}
\frac {A^{\dag}(n)}{\sqrt{n}} z^n} e^{-\alpha\cdot\sum_{n>0} \frac 
{a(n)}{\sqrt{n}}
z^{-n}}
\end{equation}
\begin{equation}
\label{eq6.25b}
[q_i,\pi_j(0)] = i\delta_{ij},\quad [L_B(0),U(\alpha,z)] = (z\partial_z +
\frac {\alpha^2}{2})U(\alpha,z)
\end{equation}
\vspace{-0.1in}
\begin{equation}
\label{eq6.25c}
U(\alpha,0)|0\> = |\alpha\>,\quad (L_B(0) - \frac {\alpha^2}{2})|\alpha\> = 0
\end{equation}
where ${\alpha}$ is an arbitrary $D$-vector. The  reduced ``Hamiltonian" $L_B(0)$ of this
system is given in Eq.(3.8d). The operator products of these more general
constructions satisfy
\setcounter{subsection}{26}
\setcounter{equation}{0}
\begin{eqnarray}
U(\alpha,z)U(\beta,\omega) &=& (z-\omega)^{\alpha\cdot\beta} e^{i(\alpha+\beta)\cdot q} 
   z^{\alpha\cdot\pi(0)} \omega^{\beta\cdot\pi(0)} \nonumber \\
& & \times e^{\sum_{n>0} \frac{A^{\dag}(n)}{\sqrt{n}} \cdot 
   (\alpha z^n + \beta \omega^n)} \Omega(-\alpha,z)\Omega(-\beta,\omega) \label{eq6.26a} \\
& & \nonumber \\
&\simeq & U(\beta,\omega)U(\alpha,z)(-1)^{\alpha\cdot\beta} 
   R(\alpha,z;\beta,\omega)  \label{eq6.26b}
\end{eqnarray}
\vspace{-0.2in}
\begin{equation}
\label{eq6.26c}
\Omega(\alpha,z) \equiv e^{\alpha\cdot\sum_{n>0} \frac {a(n)}{\sqrt{n}} z^{-n}}
\end{equation}
\begin{equation}
\label{eq6.26d}
R(\alpha,z;\beta,\omega) \equiv
\Omega(\alpha,z)\Omega(\beta,\omega)\Omega(-\alpha,\omega)\Omega(-\beta,\omega)
\end{equation}
and the commutants $\{R(\alpha,z; \beta,w)\}$ satisfy
\setcounter{subsection}{27}
\setcounter{equation}{0}
\begin{equation}
\label{eq6.27a}
[R(\alpha,z;\beta,\omega),A_i{}^{\dag}(m)] = 0,\quad R(\alpha,z;\beta 
\omega)|\gamma\> = |\gamma\>
\end{equation}
\begin{equation}
\label{eq6.27b}
R(\alpha,z;\beta,\omega)A_W{}^{\dag}|\gamma\> = A_W{}^{\dag}|\gamma\>
\end{equation}
in parallel with the quasi-fermionic example of the previous subsection.

The relations (6.26),(6.27) guarantee the free-algebraic vertex operator
construction of many complex chiral quasi-fermions
\setcounter{subsection}{28}
\setcounter{equation}{0}
\begin{equation}
\label{eq6.28a}
{\bar \psi}_i(z) = U(e_i,z){\bar K}_i,\quad \psi_i(z) = K_iU(-e_i,z)
\end{equation}
\begin{equation}
\label{eq6.28b}
(e_i)_j = \delta_{ij},\quad i,j = 1\dots D
\end{equation}
\vspace{-0.2in}
\begin{equation}
\label{eq6.28c}
{\bar K}_i = \prod_{j=1}^i e^{-i\pi\pi_j(0)},\quad K_i = \prod_{j=1}^i 
e^{i\pi\pi_j(0)}
\end{equation}
where $\bar{K},K$ are the Klein transformations used in the conventional vertex operator
construction of many complex chiral fermions [21-23]. In particular, the relations
(6.27) guarantee that, as above, these operators behave as a set of $D$ ordinary
anticommuting complex chiral fermions when acting on any of the Bose symmetric states
$A_W{}^{\dag}|\gamma\>$.

One may also consider the free-algebraic analogue of other vertex operator
constructions, for example by choosing $\alpha,\beta \in {\Delta(slg)}$ with $\alpha^2 = \beta^2 = 2$
to be roots of a simply laced Lie algebra. But it seems (see however 
Subsec. 6.4) that
this construction will not directly reproduce
the simply laced affine algebra, even on the Bose symmetric states.  The reason is
the complexity of the non-leading terms  in OPEs such as:
\setcounter{subsection}{29}
\setcounter{equation}{0}
\begin{equation}
\label{eq6.29a}
U(\alpha,z)U(-\alpha,\omega) = \frac {1}{(z-\omega)^2} + \frac 
{\alpha\cdot P(\omega)}{z-\omega} + O(z-\omega)
\end{equation}
\vspace{-0.1in}
\begin{equation}
\label{eq6.29b}
zP_i(z) = \pi_i(0) + \sum_{n>0} \sqrt{n} \left\{A_i{}^{\dag}(n)z^n + \int_0^1
dt\Omega(-\alpha t,z)a_i(n)\Omega(\alpha t,z)z^{-n}\right\}.
\end{equation}
In particular, the modes of the operators $\{P_i(z)\}$ do not satisfy an abelian current
algebra.

\setcounter{subsection}{3}
\subsection{Cuntz-algebraic factorization of the Koba-Nielsen factor}
\label{sec6.4}

As a final topic, let us consider the reduced ground state expectation value
of a product of an arbitrary number of free-algebraic vertex operators.  The zero
mode factor is  the same as usual, and so we procede by moving say the ``$A^{\dag}$ factors"
to the left until we reach the configuration:
\renewcommand{\theequation}{\thesection.\arabic{equation}}
\setcounter{equation}{29}
\begin{equation}
\label{eq6.30}
\<0| e^{\sum_{I=1}^M \alpha_{I} \cdot \sum_{n>0} \frac 
{A^{\dag}(n)}{\sqrt{n}}
z_I{}^{-n}} \Omega(-\alpha_1,z_1)\dots \Omega(-\alpha_M,z_M)|0\> = 1.
\end{equation}
Since the algebra of the $a$'s with the $A^{\dag}$'s is  also normal, our result  is
a new {\it Cuntz-algebraic factorization} of the  ordinary [10] 
{\it Koba-Nielsen factor}
\begin{equation}
\label{eq6.31}
\<0|U(\alpha_1,z_1)\dots U(\alpha_M,z_M)|0\> = \delta^D \left( \sum_{I=1}^M
\alpha_I\right) \prod_{I<J} (z_I-z_J)^{\alpha_I\cdot \alpha_J}
\end{equation}
and hence a new  Cuntz-algebraic factorization of conventional string correlators!
As a simple example the reduced ground state expectation values of the quasi-fermions
(6.28) are exactly the same as those of the corresponding set of conventional BH 
fermions. 

At first sight this result is quite surprising because, owing to the Boltzmann
statistics of the Cuntz algebra, there are many more Cuntz states
 $\{a_w{}^{\dag}| 0\>\}$
in the completeness sums of each channel than there are bosonic states in the
factorization of the conventional  bosonic string. It is not difficult however to
check that 
\begin{equation}
\label{eq6.32}
U(\alpha_i,z_i)\dots U(\alpha_M,z_M)|0\> = \sum_W 
C_W(\{\alpha\},\{z\})A_W{}^{\dag}
|\alpha_i + \dots + \alpha_M\>
\end{equation}
so that in fact only the Bose-symmetric subset ${A_W{}^{\dag}|\gamma\>}$ of the Cuntz states
are coupling in each channel of the free-algebraic factorization -- and these states
are in  one-to-one correspondence with the Bose states of the conventional bosonic
string. Moreover, the general result (6.31) defines a subspace of the reduced 
Hilbert space in which one finds effectively all the usual string constructions, including
the conventional bosonic Virasoro generators and the conventional vertex operator construction
of level one of simply laced g.

The story does not end here however because, in the free-algebraic factorization, the
rest of the Cuntz states can also couple.  As an example, the amplitudes for
scattering into the lowest non-Bose symmetric Cuntz states are proportional to: 
\begin{equation}
\label{eq6.33}
\<0| U(\alpha_1z_1)\dots 
U(\alpha_Mz_M)[a_i{}^{\dag}(m),a_j{}^{\dag}(n)]|\gamma\> \ne 0.
\end{equation}
The explicit form of these new amplitudes is beyond the scope of this paper. I
should also mention the existence of another set of free-algebraic vertex operators
using the pair $(a^{\dag}, A)$ instead of $(A^{\dag}, a)$.

I am not sanguine about the viability of such new {\it free-algebraic strings},  even when
the free-algebraic degrees of freedom are  limited to compactified dimensions. The
reason is the very large number of Cuntz states, whose partition functions
(see Subsecs. 2.5 and 3.1) would be expected to cause new singularities in
free-algebraic string loops. Similarly, formulations  of space-time canonical
ensembles are apparently ruled out by the rapid growth
\begin{equation}
\label{eq6.34}
O(e^{an}) \rightarrow O(e^{aE^2})
\end{equation}
in the number of high-mass Cuntz states.

\section*{Acknowledgements}

\medskip
For helpful discussions, I thank K.~Bardakci, O.~Ganor, C.~Helfgott, 
N.~Obers, C.~Schwartz, M.~Staudacher, C.~Thorn and F.~Wagner.

This work was supported in part by the Director,
Office of Energy Research, Office of High Energy and Nuclear Physics, 
Division of High Energy Physics of the U.S. Department of Energy under 
Contract DE-AC03-76SF00098 and in part by the National Science Foundation 
under grant PHY00-98840.

After submission of this paper, I was reminded that the large $N$ limit of $Z_N$
parafermions was studied by different methods in Ref.[25].

\vskip 0.8 in
\appendix
\section{Local reduced fields and operator products}

Although they are not used in the text, one may define ``local'' reduced fields
as usual
\renewcommand{\theequation}{\thesection.\arabic{subsection}\alph{equation}}
\setcounter{section}{1}
\setcounter{subsection}{1}
\setcounter{equation}{0}
\begin{equation}
\label{eqA.1a}
H(z) \equiv \sum_p H(p)z^{-p- 1/2},\quad J(z) \equiv \sum_m J(m)z^{-m- 1}
\end{equation}
\vspace{-0.2in}
\begin{equation}
\label{eqA.1b}
\pi(z) \equiv \sum_m \pi(m)z^{-m-1}
\end{equation}
in terms of the reduced modes $H,J$ and $\pi$ of the text. These reduced fields serve as
interpolating fields in the usual way 
\setcounter{subsection}{2}
\setcounter{equation}{0}
\begin{equation}
\label{eqA.2a}
\lim_{z \rightarrow 0} H(z)|0\> = H(-1/2)|0\>,\quad \lim_{z \rightarrow 
0} J(z)|0\> =
J(-1)|0\>
\end{equation}
\vspace{-0.1in}
\begin{equation}
\label{eqA.2b}
\lim_{z \rightarrow 0} \pi(z)|0\> = \pi(-1)|0\>
\end{equation}
and the vev's of the reduced  fields are not difficult to evaluate from
the free algebras, for example:
\setcounter{subsection}{3}
\setcounter{equation}{0}
\begin{equation}
\label{eqA.3a}
\<0|H(z)H(\omega)|0\> = \frac {1}{z-\omega},\quad \<0|\pi(z)\pi(\omega)|0\> = \frac
{1}{(z-\omega)^2}
\end{equation}
\begin{equation}
\label{eqA.3b}
\<0|J(z)J(\omega)|0\> = \frac {\hat x}{(z-\omega)^2},\quad 
\<0|J(z_1)J(z_2)J(z_3)|0\> = \frac
{-{\hat x}}{z_{12}z_{13}z_{23}}.
\end{equation}
I remind the reader (see Eq.(2.3b)) that the reduced vev's are proportional
to the corresponding traced Wightman functions at large $N$ in the unreduced theories.

The commutation relations of the reduced $\mathfrak{sl}(2)$ generators with the local reduced fields
\setcounter{subsection}{4}
\setcounter{equation}{0}
\begin{equation}
\label{eqA.4a}
[L_F(m),H(z)] = z^m(z\partial_z + \frac{1}{2}(m+1))H(z)
\end{equation}
\vspace{-0.2in}
\begin{equation}
\label{eqA.4b}
[L_F(m),J(z)] = z^m(z\partial_z + (m+1))J(z)
\end{equation}
\begin{equation}
\label{eqA.4c}
[L_B(m),\pi(z)] = z^m(z\partial_z + (m+1))\pi(z)
\end{equation}
also follow from the corresponding mode commutators of the text. These
relations and the fact that the reduced fermionic ground state is
$\mathfrak{sl}(2)$ invariant
gives the usual $\mathfrak{sl}(2)$ Ward identities
\setcounter{subsection}{5}
\setcounter{equation}{0}
\begin{equation}
\label{eqA.5a}
A(z) \equiv \<0|H(z_1)\dots H(z_M)|0\>
\end{equation}
\vspace{-0.2in}
\begin{equation}
\label{eqA.5b}
\sum_{j=1}^M z_j{}^m (z_j\partial_j + \frac{1}{2}(m+1))A(z) = 0,\quad 
|m|\le 1
\end{equation}
for the reduced fermionic vev's, and similarly for reduced vev's including
the reduced local currents and bosons.   Reduced $\mathfrak{osp}(1|2)$ Ward identities
can also be obtained for the reduced vev's of the NS sector discussed in Sec.5.  

Let us also introduce the local {\it tilde} fields
\setcounter{subsection}{6}
\setcounter{equation}{0}
\begin{equation}
\label{eqA.6a}
{\tilde H}(z) \equiv \sum_p {\tilde H}(p)z^{-p-1/2},\ {\tilde \pi}(z) 
\equiv \sum_m
{\tilde \pi}(m)z^{-m-1}
\end{equation}
\begin{equation}
\vspace{-0.3in}
\label{eqA.6b}
({\tilde H}(z) - H(z))|0\> = ({\tilde \pi}(z)-\pi(z))|0\> = 0
\end{equation}
\begin{equation}
\label{eqA.6c}
\<0|({\tilde H}(z) - H(z)) = \<0|({\tilde \pi}(z)-\pi(z)) = 0
\end{equation}
where the tilde modes are defined in Eqs.(2.4) and (3.3). The untilde
modes satisfy commutation and/or anticommutation relations with the tilde
modes, so we may compute in more or less the usual fashion the operator
products of the untilded  fields with the tilde fields.  For example,
the operator products
\setcounter{subsection}{7}
\setcounter{equation}{0}
\begin{equation}
\label{eqA.7a}
H(z){\tilde H}(\omega) = \frac {|0\>\<0|}{z-\omega} +
{\mathbf :}H(z){\tilde H}(\omega){\mathbf :}
\end{equation}
\vspace{-0.1in}
\begin{eqnarray}
\label{eqA.7b}
J(z){\tilde H}(\omega) &=& \frac {|0\>\<0|{\tilde H}(\omega) - {\tilde
H}(\omega)|0\>\<0|}{z-\omega} + {\mathbf :}J(z){\tilde H}(\omega){\mathbf 
:} \\
&=& \frac {|0\>\<0|H(\omega) - H(\omega)|0\>\<0|}{z-\omega} + {\mathbf :}J(z){\tilde H}(\omega){\mathbf 
:} 
\end{eqnarray}
\vspace{-0.1in}
\begin{equation}
\label{eqA.7d}
\pi(z){\tilde \pi}(\omega) = \frac {|0\>\<0|}{(z-\omega)^2} + 
{\mathbf :}\pi(z){\tilde
\pi}(\omega){\mathbf :}
\end{equation}
are obtained for $|z|>|w|$, where I have used the conventional fermionic and/or bosonic
normal ordering for untilde/tilde mode pairs.

Finally, the reduced fields satisfy the expected Hamiltonian equations of
motion on the cylinder $(t,\xi)$, for example
\setcounter{subsection}{8}
\setcounter{equation}{0}
\begin{equation}
\label{eqA.8a}
H(\xi,t) \equiv \sum_p H(p,t)e^{-i\xi p} = \sum_p H(p,0)e^{-i(\xi+t)p}
\end{equation}
\vspace{-0.2in}
\begin{equation}
\label{eqA.8b}
\partial_t H(p,t) = i[L_F(0),H(p,t)] = -ipH(p,t)
\end{equation}
\begin{equation}
\label{eqA.8c}
\partial_t H(\xi,t) = i[L_F(0),H(\xi,t)] = \partial_{\xi} H(\xi,t)
\end{equation}
so that the reduced fermion is a chiral field.

\vskip .5cm
%\clearpage
\addcontentsline{toc}{section}{References} 

\renewcommand{\baselinestretch}{.4}\rm
{\footnotesize

\providecommand{\href}[2]{#2}\begingroup\raggedright

\end{document}